\documentclass[journal]{IEEEtran}

\ifCLASSINFOpdf
\else
\fi

\usepackage{enumitem}
\usepackage{array}
\usepackage{times}
\usepackage{epsfig}
\usepackage{graphicx}
\usepackage{amsmath}
\usepackage{amssymb}
\usepackage{graphicx}
\usepackage{cite}
\usepackage{cases}
\usepackage{multirow}
\usepackage{verbatim}
\usepackage{amssymb}
\usepackage{CJK}
\usepackage{dblfloatfix}   
\usepackage{placeins}
\usepackage{algorithm}
\usepackage{algorithmicx}
\usepackage{algpseudocode}
\usepackage{color}
\usepackage[dvipsnames]{xcolor}
\usepackage{bm}
\usepackage{booktabs}
\usepackage[colorlinks,linkcolor=red]{hyperref}
\usepackage{cuted}

\begin{document}
\bstctlcite{BSTcontrol}

\title{FFDT: Sim-to-Real Domain Transfer Method for Far-Field Ground Penetrating Radar}


\author{Chi Wang, Hai-Han Sun,~\IEEEmembership{Senior Member,~IEEE}
\thanks{ 
}
\thanks{The authors are with the Department of Electrical and Computer Engineering, University of Wisconsin–Madison, USA, 53706 (e-mail: cwang788@wisc.edu, haihan.sun@wisc.edu).}
}

\markboth{}%
{Shell \MakeLowercase{\textit{et al.}}: Bare Demo of IEEEtran.cls for IEEE Journals}

\maketitle

\begin{abstract}
The scarcity of real-world ground penetrating radar (RealGPR) data is a major barrier to the development of deep learning (DL) methods for GPR data processing. Although simulated GPR (SimGPR) datasets can be used to train DL models, differences in hardware configuration and system parameters between SimGPR and RealGPR create a sim-to-real domain gap that leads to poor generalization of DL models to RealGPR measurements. To address this gap, we present a physics-based far-field GPR domain transfer (FFDT) method to transfer far-field SimGPR data to the corresponding RealGPR-domain data in the same scenario. In FFDT, SimGPR and RealGPR are modeled using angle-dependent transfer functions that capture three-dimensional antenna radiation characteristics. These transfer functions are then combined with ray tracing to enable the sim-to-real data transfer. Extensive studies across diverse objects and soil conditions demonstrate that FFDT-transferred data closely match RealGPR ground truth. To further demonstrate its value for RealGPR-domain dataset generation, FFDT is implemented to build a RealGPR-domain dataset for training a DL model for clutter removal. The model achieves robust performance on RealGPR measurements. With one-time SimGPR and RealGPR system calibration, FFDT enables RealGPR-domain dataset generation without requiring task-specific RealGPR field measurements for DL model training or fine-tuning, thereby addressing the RealGPR data scarcity bottleneck and supporting reliable development and deployment of DL methods for real-world GPR applications.

\end{abstract}


\IEEEpeerreviewmaketitle

\section{Introduction}
\IEEEPARstart {G}{round} penetrating radar (GPR) is a non-invasive subsurface sensing technology that uses electromagnetic waves to detect subsurface objects. It has broad applications in diverse disciplines, including agriculture, civil engineering, defense, and planetary exploration~\cite{ag,civil,env,defense,planetary}. To meet the growing demand for real-time operation with higher detection and characterization accuracy,  artificial intelligence (AI), particularly deep learning (DL) methods, has been actively investigated. By learning the relationships between GPR data and task-specific  outputs from large datasets in a one-time training stage, DL models can perform data processing and generate results orders of magnitude faster and with enhanced accuracy compared to conventional methods. Despite recent progress, the development of DL-based GPR methods remains fundamentally constrained by the \textbf{scarcity of real-world GPR (RealGPR) data}. As data-driven methods, DL models rely heavily on training datasets that are high-quality, diverse, and well aligned with the real-world deployment domain (i.e., RealGPR domain). However, collecting RealGPR data is costly and time-consuming, and accurate ground-truth labeling is often difficult. As a result, full-wave electromagnetic simulation is commonly used to generate large-scale training datasets based on simulated GPR (SimGPR) systems. However, even for the same sensing scenario, SimGPR and RealGPR signals can differ substantially due to differences in hardware configuration, antenna radiation characteristics, and other system parameters between the two systems. This  SimGPR-RealGPR domain mismatch (i.e., the sim-to-real domain gap)  often causes DL models trained on SimGPR data to generalize poorly to RealGPR measurements \cite{UnsupervisedLearningReference1,Wavelet-Gan}. \par

Three main learning-based strategies have been adopted to reduce the sim-to-real domain gap. The first strategy is supervised transfer learning, in which a DL model is first trained on a large-scale SimGPR dataset to learn generic scattering physics and then fine-tuned using a small amount of labeled RealGPR data to adapt to the RealGPR domain \cite{env, 3Dimage,2Dimage, tree}. The second strategy is to construct hybrid datasets for model training, either by integrating features from both SimGPR and RealGPR domains \cite{GPRInvNet, clutter1, Wavelet-Gan} or by directly mixing SimGPR and RealGPR datasets \cite{mix1, mix2, mix3}. The third strategy employs unsupervised domain adaptation \cite{UnsupervisedLearningReference1,UnsupervisedLearningReference2,UnsupervisedLearningReference3,UnsupervisedLearningReference4}, which  learns domain-invariant feature representations from labeled SimGPR data and unlabeled task-specific RealGPR data through adversarial training to reduce the domain gap. Although these strategies can improve DL model performance on the RealGPR domain, they still require task-specific RealGPR measurements for training, and their effectiveness depends strongly on data quality, diversity and, for supervised learning, label accuracy. Consequently, RealGPR data acquisition remains a key bottleneck for the reliable development and real-world deployment of DL-based GPR methods. \par

\textcolor{black}{Physics-based GPR calibration methods have been developed to reduce waveform discrepancies between SimGPR and RealGPR. The wavelet cross-convolution method reduces the influence of SimGPR-RealGPR wavelet differences by cross-convolving simulated and measured responses~\cite{FWI_crossconv1,FWI_crossconv2,FWI_crossconv3}. Wavelet deconvolution methods, including iterative wavelet estimation and sparse blind deconvolution~\cite{FWI_deconv,SBD1,SBD2}, characterize an effective source wavelet and use deconvolution to reduce the waveform mismatch. In addition, GPR system responses can be characterized using transfer functions for accurate forward modeling and inversion ~\cite{Lambot1,lambot3}. Although these methods differ in formulation, they generally characterize the source or system response using an effective wavelet or angle-independent transfer function. Such representations do not explicitly account for the angle-dependent transmitting and receiving characteristics of practical GPR antennas and could therefore leave residual sim-to-real discrepancies when reflections arrive from different directions.}\par 

\textcolor{black}{In this paper, we address the sim-to-real domain gap by developing a novel GPR domain transfer method, referred to as FFDT, for (quasi-)monostatic, co-polarized GPR systems operating in the far field.} The objective of FFDT is to transfer SimGPR signals to the signals that would be acquired by a specified RealGPR system in the same scenario. By replacing task-specific RealGPR training data collection with physics-based domain transfer, FFDT offers a feasible pathway to alleviating the RealGPR data scarcity barrier. \par

The main contributions of this work are summarized as follows:
\begin{itemize}
\item A far-field (quasi-)monostatic GPR model is presented by extending the antenna model in~\cite{Lambot1}.  Angle-dependent transfer functions are introduced to incorporate three-dimensional antenna radiation characteristics,  which is critical for the FFDT method. 
\item A physics-based FFDT method is developed that combines SimGPR and RealGPR transfer functions with ray tracing to transfer signals from the SimGPR domain to the RealGPR domain. 
\item FFDT performance is extensively evaluated across different object geometries, material properties, soil heterogeneity, surface roughness, and GPR systems. The FFDT-transferred data closely match RealGPR data, with differences below $-$30 dB in most cases. 
\item A primary application of FFDT is the generation of RealGPR-domain datasets from simulations for training DL models. Using an experimental clutter-removal task and a numerical detection and classification task, it is shown that a model trained on FFDT-transferred data achieves substantially improved performance on RealGPR radargrams compared with the same model trained on SimGPR data. These results demonstrate that FFDT enables RealGPR-domain dataset generation without requiring \textcolor{black}{task-specific} field measurements.
\end{itemize}

The remainder of this paper is organized as follows. Section~\ref{sec2} presents the FFDT method and uses a case study to demonstrate its implementation process and effectiveness. Section~\ref{sec3} evaluates FFDT performance across diverse object and subsurface scenarios, and discusses the far-field criterion. Section~\ref{sec4} presents three experimental studies and demonstrates the effectiveness of FFDT in generating RealGPR-domain datasets for DL model training.  \textcolor{black}{Section~\ref{sec5} discusses the applicability and limitations of FFDT in practical implementation. Section~\ref{sec6} compares FFDT with GPR calibration methods and learning-based domain adaptation methods.} The conclusions are drawn in Section~\ref{sec7}.

\section{Methodology}\label{sec2}

In this section, we first present a frequency-domain model of far-field monostatic GPR systems formulated using angle-dependent transfer functions, and then present a novel far-field domain transfer (FFDT) method that combines system transfer functions with ray tracing to transfer far-field GPR data from the SimGPR domain to the RealGPR domain. The GPR system model extends the antenna model in \cite{Lambot1} by incorporating three-dimensional antenna radiation characteristics, which are critical to the FFDT method.

\subsection{Far-Field Monostatic GPR System Model} \label{GPRmodel}

\begin{figure}[t]
  \centering
  \includegraphics[width=0.65\linewidth]{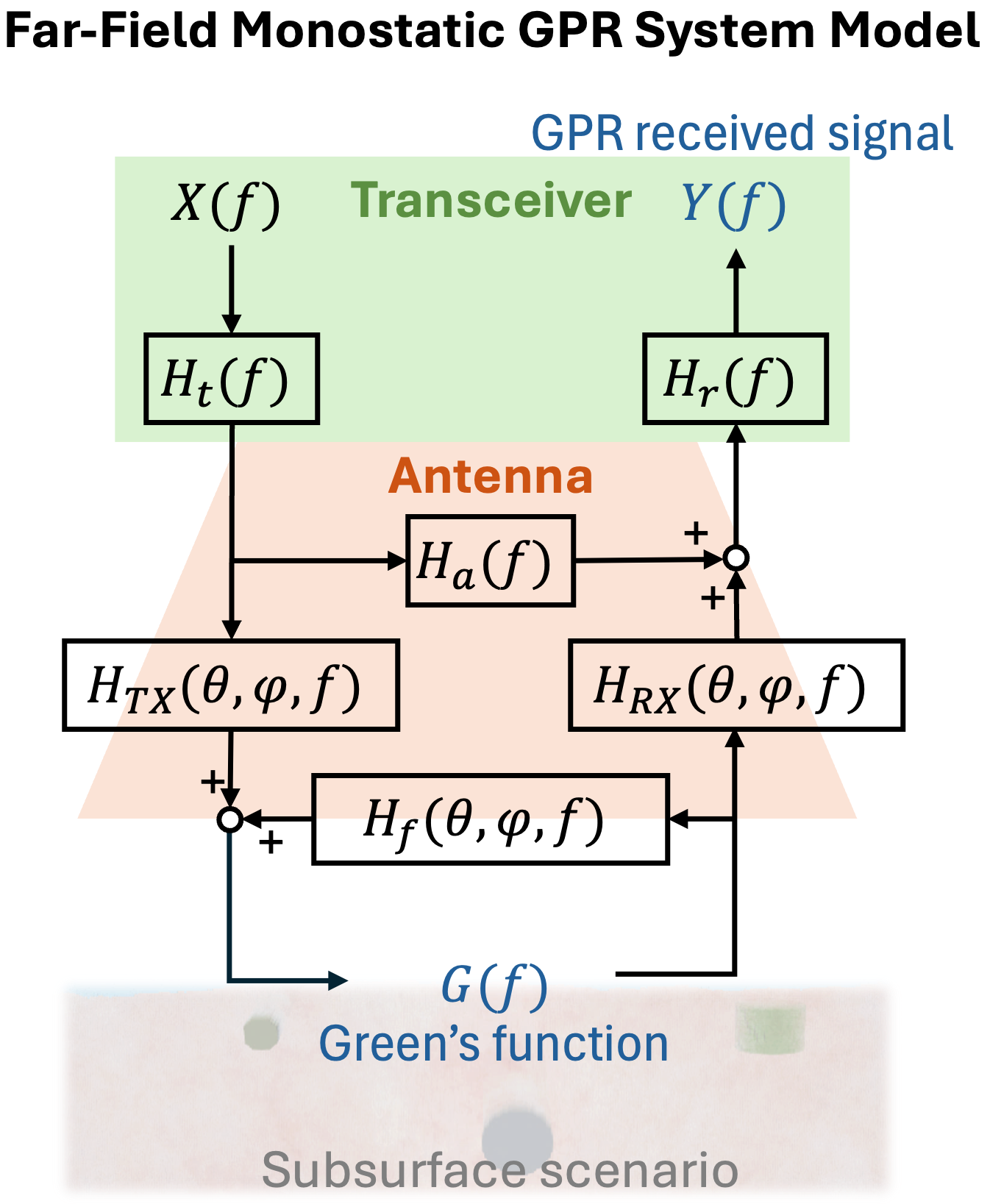}
  \vspace{-3pt}
  \caption{Far-field monostatic GPR system model in the frequency domain.  }
  \vspace{-6pt}
  \label{fig:GPRmodel}
\end{figure}

As shown in Fig.~\ref{fig:GPRmodel}, the monostatic GPR system, which includes a transceiver and an antenna functioning as both transmitter (TX) and receiver (RX), is modeled in the frequency domain. In the far field, the antenna can be approximated as a point source and receiver. At a given frequency $f$, $X(f)$ and $Y(f)$ are the transmitted and received signals, respectively. $H_t(f)$ and $H_r(f)$ are the transmitting and receiving transfer functions of the transceiver, respectively. $H_a(f)$ represents the antenna reflection coefficient in free space. $H_{\mathrm{TX}}(\theta,\phi,f)$ and $H_{\mathrm{RX}}(\theta,\phi,f)$ are the transmitting and receiving transfer functions of the antenna in the spatial direction $(\theta,\phi)$. These two functions simultaneously account for the antenna’s radiation intensity and phase delay in three-dimensional space. The positive feedback function $H_f(\theta,\phi,f)$ accounts for multiple reflections between the antenna and the sensing environment along the wave propagation path. $G(f)$ represents the Green’s function of the sensing environment, describing the scattered electric field at the RX in response to a unit-strength electric source at the TX. This function is independent of the GPR system. 

It is noted that the monostatic GPR system model shown in Fig.~\ref{fig:GPRmodel} is also applicable to the quasi-monostatic configuration, in which two antennas serving as TX and RX are placed in close proximity so that the bistatic angle is small and the propagation paths are nearly identical to those of a monostatic system. In this case, $H_a(f)$ represents the direct TX-RX coupling, while the remainder of the model remains unchanged.

Using this model, $Y(f)$ is related to $G(f)$ by

\begin{equation}
\begin{aligned}
Y(f) =\;& X(f)\,H_t(f)\,H_r(f) \\
&\Bigg[
H_a(f)
+ \frac{
H_{\mathrm{TX}}(\theta,\phi,f)\,
H_{\mathrm{RX}}(\theta,\phi,f)\,
G(f)
}{
1 - H_f(\theta,\phi,f)\,G(f)
}
\Bigg].
\end{aligned}
\end{equation}
The transceiver’s system transfer functions $X(f)H_t(f)H_r(f)$ can typically be calibrated using a standard calibration kit, or a known transmission line or load. Therefore, we assume that they are known and are removed from the received signal, and will not include them in the following analysis. The remaining part of Eq.~(1) that depends solely on the antenna transfer functions is re-defined as $Y(f)$:
\begin{equation}
\begin{aligned}
Y(f) =\;& H_a(f) + \frac{
H_{\mathrm{TX}}(\theta,\phi,f)\,
H_{\mathrm{RX}}(\theta,\phi,f)\,
G(f)
}{
1 - H_f(\theta,\phi,f)\,G(f)
}.
\end{aligned}
\label{eq:antenna_only_model}
\end{equation}

\begin{figure*}[t]
\begin{equation}
\begin{aligned}
\begin{bmatrix}
Y_1(f) - H_a(f) \\
Y_2(f) - H_a(f) \\
\vdots \\
Y_N(f) - H_a(f)
\end{bmatrix}
=\;&
\begin{bmatrix}
G_1(f) & G_1(f)\!\left[Y_1(f) - H_a(f)\right] \\
G_2(f) & G_2(f)\!\left[Y_2(f) - H_a(f)\right] \\
\vdots & \vdots \\
G_N(f) & G_N(f)\!\left[Y_N(f) - H_a(f)\right]
\end{bmatrix}
\begin{bmatrix}
H_{\mathrm{TX}}(\theta,\phi,f)\,H_{\mathrm{RX}}(\theta,\phi,f) \\
H_f(\theta,\phi,f)
\end{bmatrix}
\end{aligned}
\label{eq:matrix_calibration}
\end{equation}
\end{figure*}

The antenna reflection coefficient $H_a(f)$ can be obtained by performing a GPR measurement in free space, where $G(f)=0$. The remaining combined functions $H_{\mathrm{TX}}(\theta,\phi,f)H_{\mathrm{RX}}(\theta,\phi,f)$ and feedback function $H_f(\theta,\phi,f)$ can be determined by performing GPR measurements over an electrically large metallic plate considered as a perfect electric conductor (PEC), for which the Green’s function is known, at the transmitting/receiving angle $(\theta,\phi)$ at two or more different distances. These measurements yield at least two independent equations to derive $H_{\mathrm{TX}}(\theta,\phi,f)H_{\mathrm{RX}}(\theta,\phi,f)$ and $H_f(\theta,\phi,f)$.  

In this work, to improve numerical stability when solving for these transfer functions over a broad frequency band, the approach in~\cite{Lambot2} is adopted. GPR measurements over an electrically large PEC plate at each $(\theta,\phi)$ at more than three different distances are used to form an overdetermined system of equations of Eq.~\eqref{eq:antenna_only_model}. This system of equations can be reformulated in the matrix form shown in Eq.~\eqref{eq:matrix_calibration} to solve for $H_{\mathrm{TX}}(\theta,\phi,f)H_{\mathrm{RX}}(\theta,\phi,f)$ and  $H_f(\theta,\phi,f)$.   In this formulation, $H_a(f)$ can be obtained either by performing free-space GPR measurement, or by treating $H_a(f)$ as an additional unknown and modifying Eq.~\eqref{eq:matrix_calibration} accordingly. The derived transfer functions enable GPR domain transfer between  $Y(f)$ to $G(f)$ in either direction.

For each GPR system, the transfer-function calibration is a one-time process. Once obtained, the calibrated transfer functions can be used across different sensing environments. This is because the far-field antenna radiation characteristics are intrinsic to the GPR antennas and are independent of the specific sensing environment.

\subsection{Far-Field GPR Domain Transfer (FFDT) Method}

\begin{figure}[tb]
	\begin{center}			\includegraphics[width=1\linewidth]{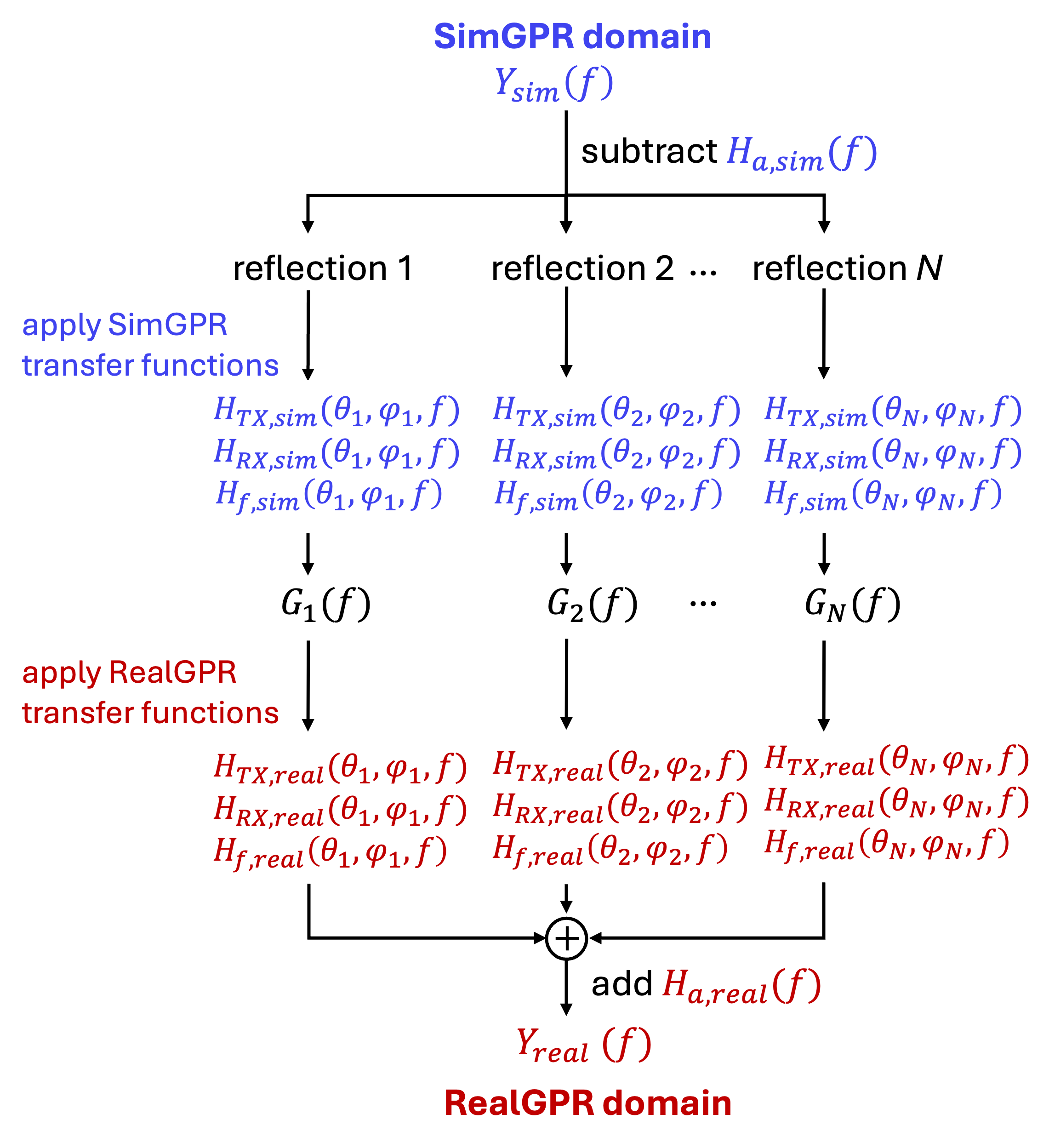}
    \vspace{-6pt}
	\end{center}
	\caption{Illustration of the far-field GPR domain transfer (FFDT) method.}
    \vspace{-0pt}
	\label{fig:FFDT}
\end{figure}

The goal of the FFDT method is to transfer received signals from a SimGPR implemented in a full-wave electromagnetic solver to the signals that would be received by a RealGPR in the same scenario. The workflow of the FFDT method is shown in Fig.~\ref{fig:FFDT}. The received signals of the two systems are denoted as $Y_{\mathrm{sim}}$ and $Y_{\mathrm{real}}$. The SimGPR system has the transfer functions $H_{a,\mathrm{sim}}$, $H_{\mathrm{TX,sim}}$,  $H_{\mathrm{RX,sim}}$, and $H_{f,\mathrm{sim}}$, while the RealGPR system has the transfer functions $H_{a,\mathrm{real}}$, $H_{\mathrm{TX,real}}$, $H_{\mathrm{RX,real}}$, and $H_{f,\mathrm{real}}$.
At each GPR survey position, the SimGPR signal $Y_{\mathrm{sim}}(f)$ includes $H_{a,\mathrm{sim}}(f)$, which can be directly subtracted, and multiple reflections from the surface and subsurface objects (reflection 1 to $N$). Since these reflections are usually captured at different transmitting/receiving angles, transfer functions at different angles should be applied to convert each reflection to the corresponding Green’s function. Therefore, the multiple reflections are first separated. Ray tracing and Snell's law are then used to estimate the transmitting/receiving angles $(\theta_n,\phi_n)$ for reflection $n$. The transfer functions  at the corresponding $(\theta_n,\phi_n)$ are applied to transfer each reflected signal to the Green’s function $G(f)$. The resulting $G(f)$ is then further transferred to the RealGPR domain by applying the RealGPR transfer functions at $(\theta_n,\phi_n)$ in Eq.~(2). The transferred signals for all the reflections are then combined, with $H_{a,\mathrm{real}}(f)$ added, to form the received signal $Y_{\mathrm{real}}(f)$ in the RealGPR domain.

The overall process can be summarized in four steps: reflection separation, angle estimation, signal transfer, and combination. It is performed in the frequency domain and can be conducted across any desired sub-bands within the overlapping frequency range of both GPR systems. Applying this process transfers SimGPR signals to RealGPR signals in the frequency domain, which can be further transformed to the time domain using the inverse Fourier transform.

\begin{figure}[tb]
	\begin{center}
    \includegraphics[width=0.9\linewidth]{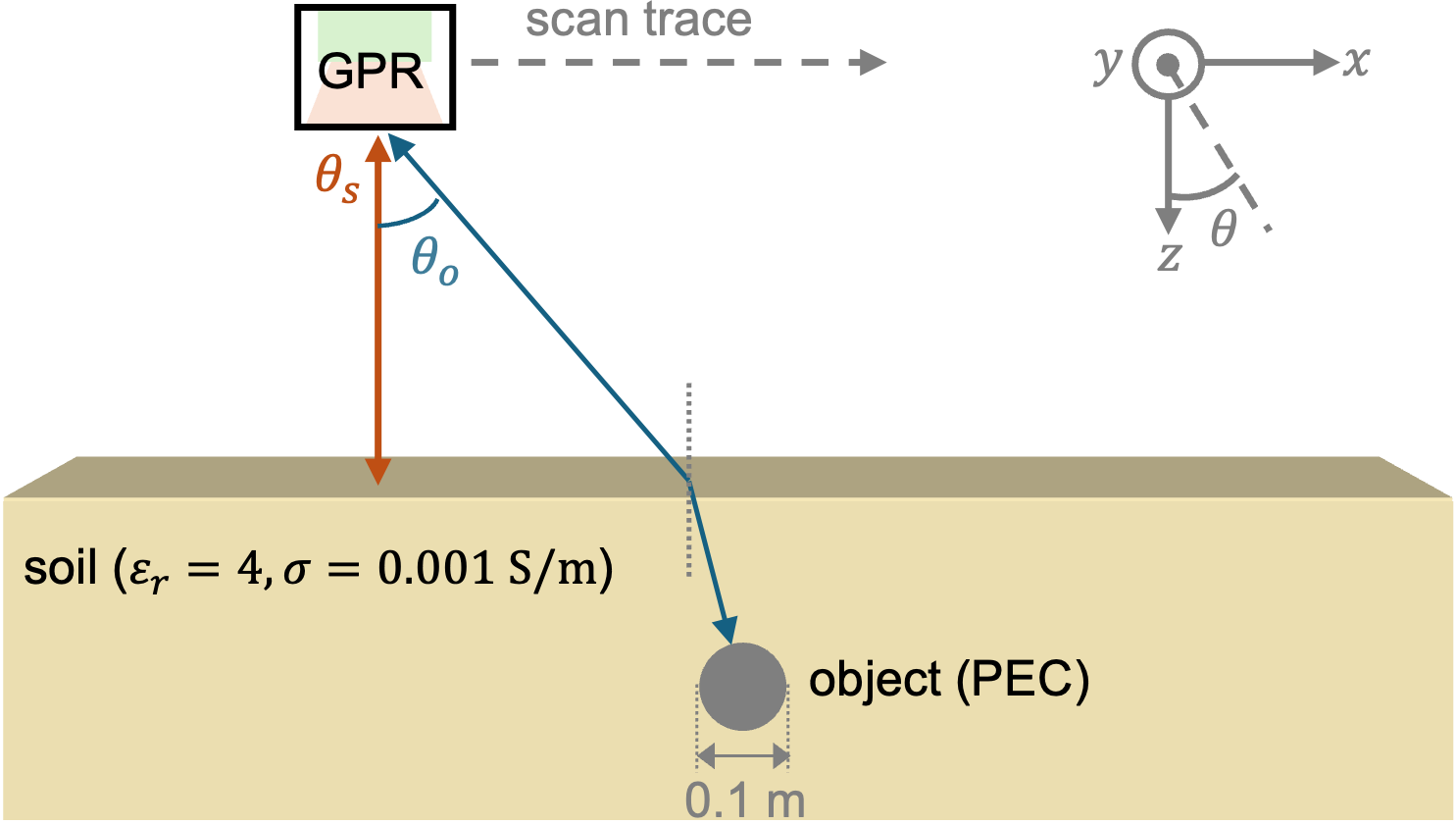}
    \vspace{-3pt}
	\end{center}
	\caption{Simulation scenario of a case study. The SimGPR and RealGPR systems perform B-scan measurements to detect a cylindrical PEC pipe in soil. The antenna-to-surface height is 0.6~m. The PEC pipe is buried at a depth of 0.2~m.}
    \vspace{-6pt}
	\label{fig:case scenario}
\end{figure}

\begin{figure*}[b]
	\centering
	\centerline{\includegraphics[width = 1\linewidth]{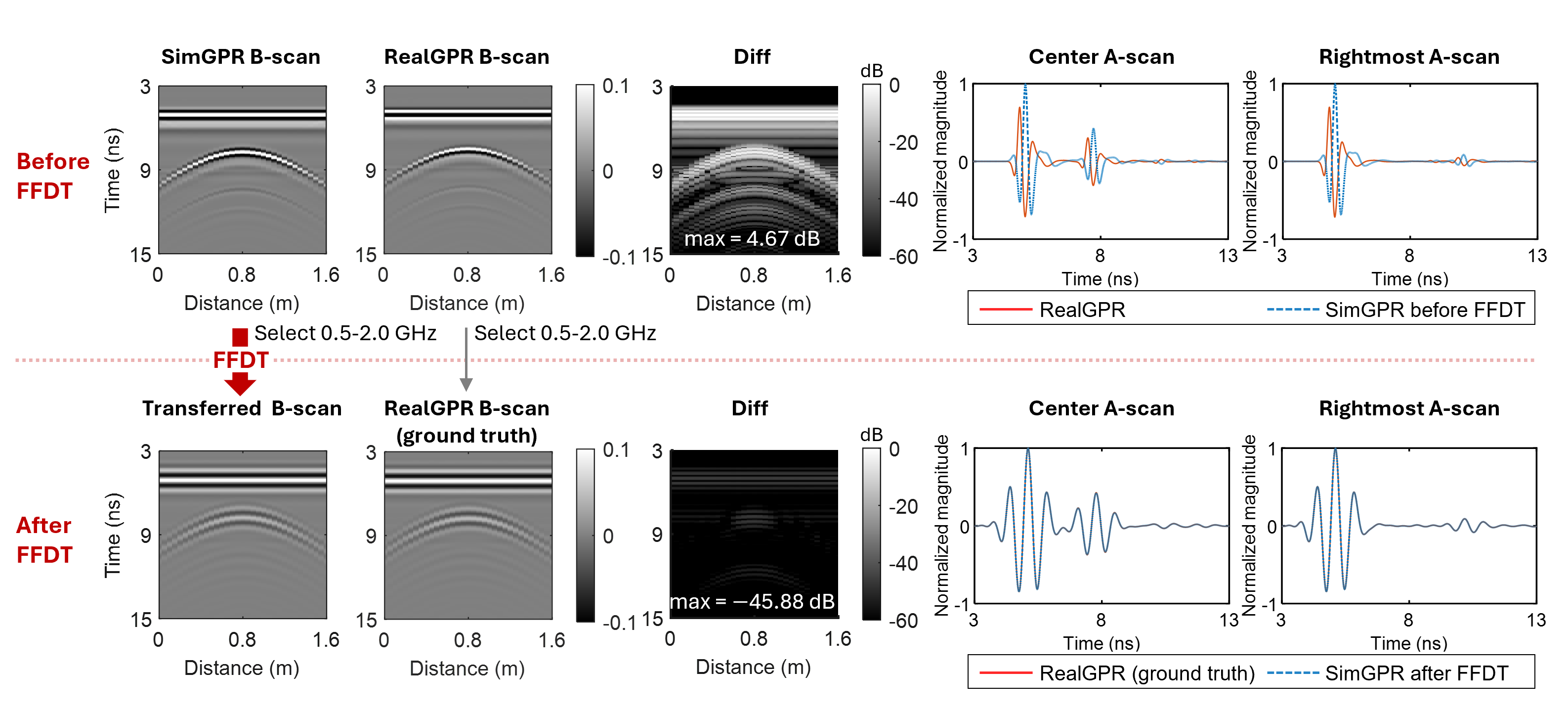}}
    \vspace{-8pt}
	\caption{Comparison of the SimGPR B-scan, FFDT-transferred B-scan, and RealGPR B-scan for the scenario in Fig.~\ref{fig:case scenario}. The transferred B-scan closely matches the RealGPR B-scan, with the maximum difference between the normalized B-scans of $-$44.17~dB. }
    \vspace{-10pt}
	\label{fig:comparison}
\end{figure*}

\subsection{Case Study} \label{sec:CaseStudy}
A case study is presented to demonstrate the implementation workflow of the FFDT method and validate its efficacy. The 2D scenario is shown in Fig.~\ref{fig:case scenario}, where two different GPR systems (SimGPR and RealGPR) are used to perform B-scan acquisitions to detect a cylindrical PEC pipe (object) in soil. The pipe has a diameter of 0.1~m and is buried at a depth of 0.2~m. The soil is modeled with a relative permittivity of 4 and a conductivity of 0.001~S/m. During B-scan acquisition, both systems have antenna-to-surface height of 0.6~m, and traces are collected over a lateral span of 1.6~m with a step size of 0.05~m. The simulation is performed using GPU-accelerated open-source software gprMax \cite{gprmax1,gprmax2}.

The RealGPR and SimGPR used for this case study have different excitation waveforms and system configurations. The RealGPR is the gprMax-modeled GSSI 1.5-GHz system \cite{GSSI1,GSSI2}. It has two bowtie antennas with TX and RX separated by 0.06~m, backed by a complex PEC enclosure filled with absorbers. The excitation waveform is the 1.5-GHz center frequency Gaussian waveform. The SimGPR is a simplified GPR system with a Hertzian dipole as TX and a field probe as the RX, separated by 0.06~m, backed by a simplified PEC box filled with absorbers. The box is constructed to maintain a similar feedback function as the RealGPR. The excitation waveform is 1.5-GHz center frequency Ricker waveform (the negative, normalized second derivative of a Gaussian waveform). 

\begin{figure}[tb]
	\centering
    \centerline{\includegraphics[width=1\linewidth]{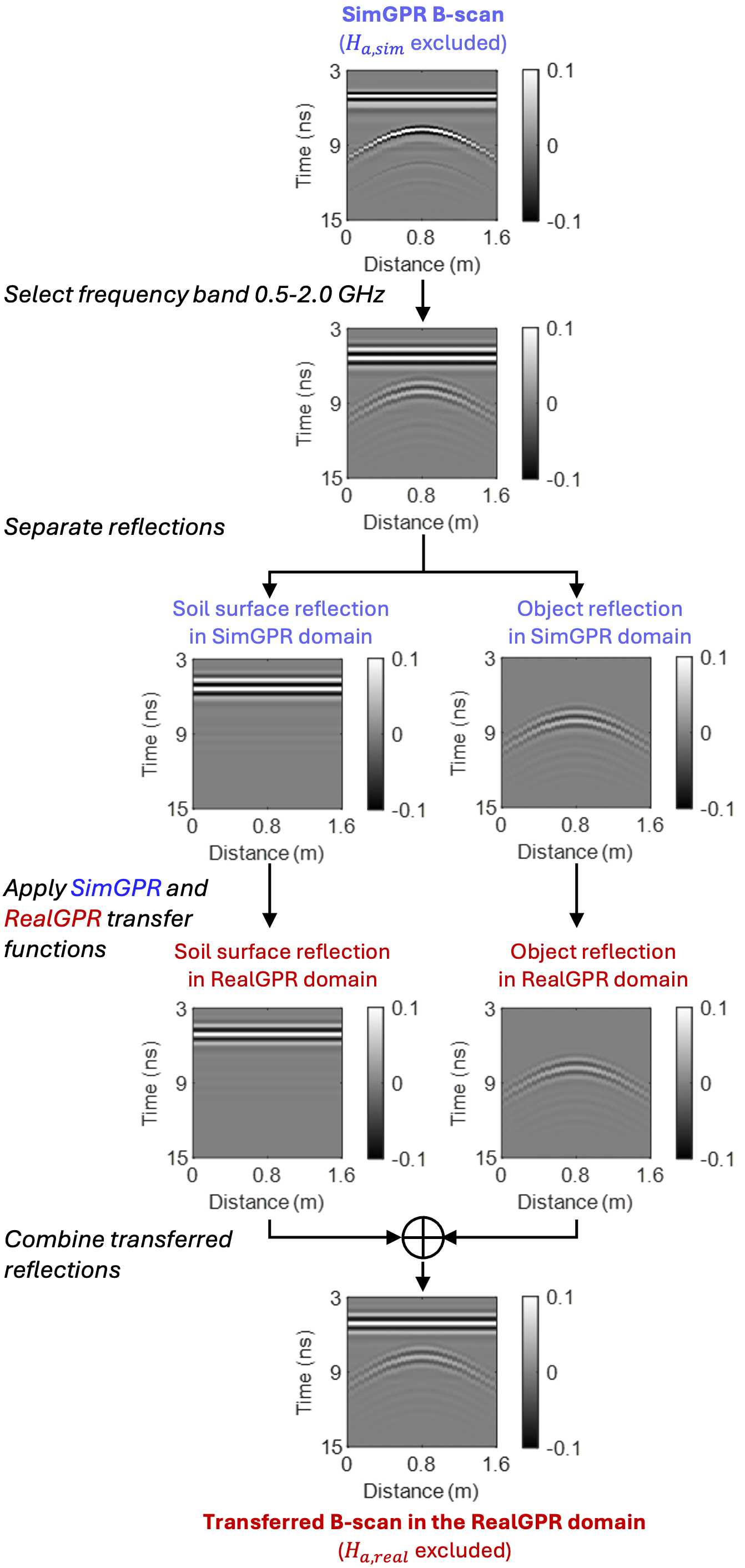}}
    \vspace{-8pt}
	\caption{Illustration of the FFDT workflow using the case study.}
    \vspace{-10pt}
	\label{fig:FFDT workflow}
\end{figure}

Since SimGPR and RealGPR have different TX/RX configurations and excitation waveforms, their transfer functions are different. As a result, their signals collected in the same scenario can differ substantially. Their B-scans collected in the scenario in Fig.~\ref{fig:case scenario} after subtracting $H_a$ are shown in the top row of Fig.~\ref{fig:comparison}. We compute the difference (Diff) between the normalized SimGPR and RealGPR B-scans as
\begin{equation}
\mathrm{Diff}
=
20 \log_{10}
\left|
\mathrm{Bscan_{n,RealGPR}}
-
\mathrm{Bscan_{n,SimGPR}}
\right|,
\end{equation}
where the subscript $\mathrm{n}$ of $\mathrm{Bscan_{n,...}}$ denotes that the B-scan is normalized to $[-1,1]$  using the maximum absolute value across the two B-scans under comparison. As shown in Fig.~\ref{fig:comparison}, the maximum Diff between the normalized SimGPR and RealGPR B-scans is 4.67~dB, demonstrating the substantial difference between the two B-scans.

FFDT is implemented to transfer signals from the SimGPR domain to the RealGPR domain. Since $H_a$ is usually considered as clutter and can be easily subtracted from far-field GPR data, all the data considered in this paper exclude $H_a$.  

\textbf{Derive GPR system transfer functions.} Since the scenario is two-dimensional, the SimGPR and RealGPR transfer functions only need to be obtained as a function of $\theta$. Following the process described in Section~\ref{GPRmodel}, for each GPR, simulations over a large PEC plate are performed for $\theta$ from $-80^\circ$ to $80^\circ$ in 2$^\circ$ increments. For each angle, measurements are taken at 16 antenna-to-plate distances ranging from 0.6~m to 0.9~m with a step size of 0.02~m. These data are then used to derive the system transfer functions using Eq.~\ref{eq:matrix_calibration}. 

\textbf{FFDT Implementation.} Fig.~\ref{fig:FFDT workflow} illustrates the FFDT implementation process. As noted earlier, $H_a$ is typically treated as noise and can be readily subtracted. Thus, the objective of the FFDT is to map the SimGPR-domain data with $H_a$ removed to the RealGPR domain. Since FFDT is implemented in the frequency domain, the time-domain signal output by gprMax is first transformed to the frequency domain using the Fourier transform, and the signal in the desired frequency band 0.5~GHz to 2.0~GHz is extracted. (For visualization, all the frequency-domain results in this work are converted back to the time domain by applying a Kaiser window ($\beta$=5) followed by the inverse Fourier transform.)

At each survey position, the received signal contains reflections from the soil surface and the buried object. Since these two reflections are captured at different transmitting/receiving angles $\theta$ (Fig.~\ref{fig:case scenario}), angle-dependent transfer functions should be applied to perform the sim-to-real transfer. Therefore, the two reflections are first separated. This separation is achieved by first simulating a soil-only case to obtain the soil-surface reflection, and then subtracting it from the full B-scan to obtain the object reflection. 

For each A-scan at each survey position, ray tracing and Snell’s law are used to estimate the $\theta_s$ and $\theta_o$ for the soil-surface and object reflections, respectively. The corresponding SimGPR and RealGPR transfer functions are applied to map each reflection to the RealGPR domain. The transferred components are subsequently superimposed to form the transferred A-scan. This process is repeated for all A-scans, and the transferred A-scans are compiled into the final transferred B-scan.

The results after applying FFDT are shown in the bottom row of Fig.~\ref{fig:comparison}. The transferred B-scan is almost identical to the RealGPR B-scan from 0.5 GHz to 2.0 GHz, with the maximum Diff reduced from 4.67~dB to $-$45.88~dB. The significant Diff reduction of more than  50~dB demonstrates the effectiveness of the FFDT method in performing the sim-to-real domain transfer.

\section{Numerical Study}\label{sec3}

Numerical studies are conducted to evaluate the FFDT performance across diverse scenarios. Specifically, Section~\ref{DiffObj} presents the FFDT performance for objects with different geometries and dielectric properties; Section~\ref{DiffSoil} examines the FFDT performance under different soil heterogeneity, surface roughness, and layered media; and Section~\ref{FarFieldCondition} discusses the far-field criterion under which FFDT can be effectively applied. \par

\begin{figure*}[!t]
	\centering
	\centerline{\includegraphics[width=1\linewidth]{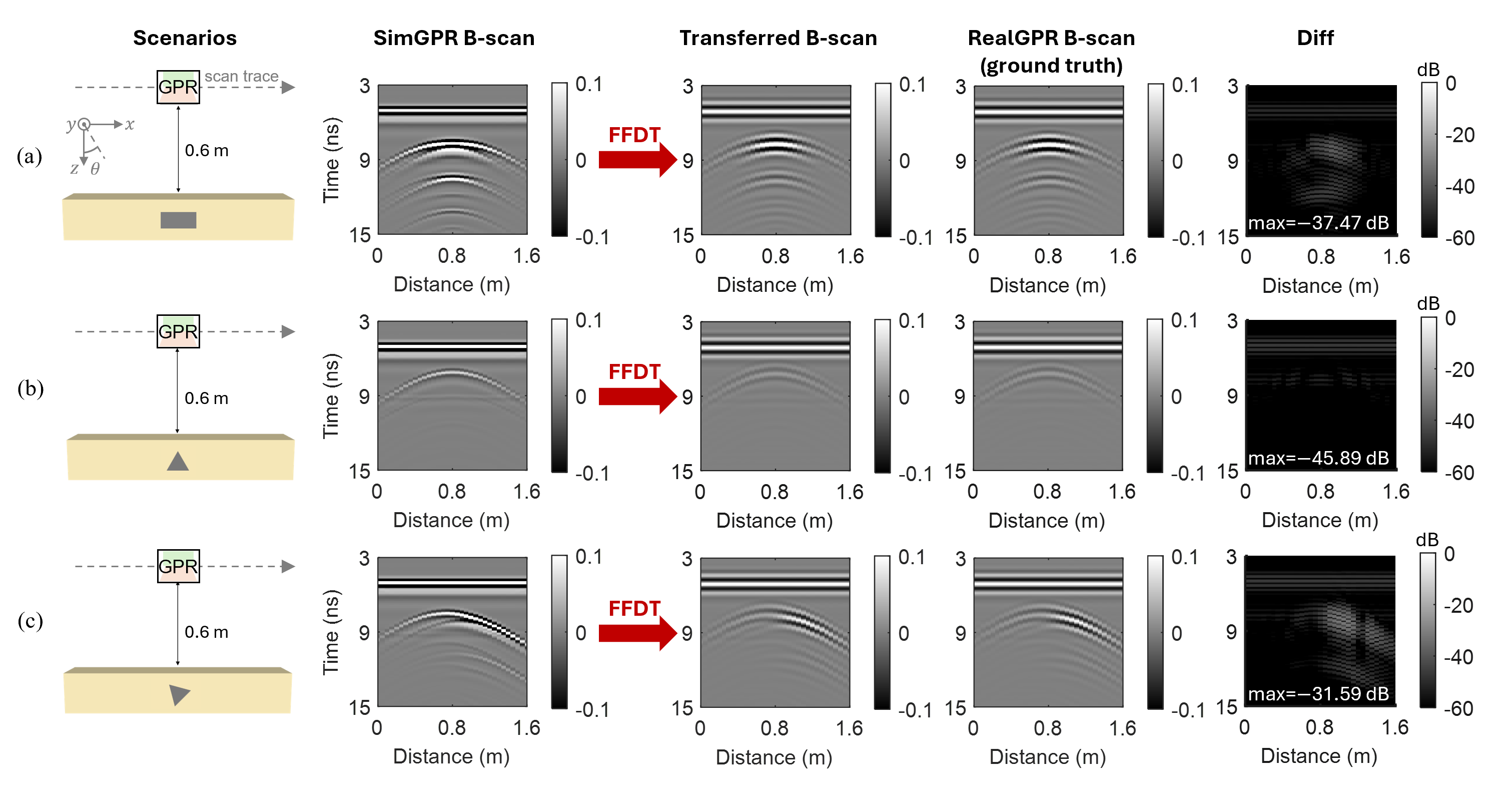}}
    \vspace{-4pt}
    \caption{Comparison of FFDT performance for PEC objects with different geometries: (a) a PEC rectangular prism with a width of 0.2~m and a height of 0.1~m. (b)  a PEC 	equilateral triangular prism with a side length of 0.173~m at 0$^\circ$ rotation. (c) the same PEC equilateral triangular prism at 45$^\circ$ rotation.}
	\label{fig:PEC objects comparison}
\end{figure*}

\subsection{FFDT Performance with Different Objects } \label{DiffObj}

\begin{table*}[tb]
\centering
\caption{Maximum Diff Between RealGPR B-Scan and SimGPR B-scan Before and After FFDT in Different Object Cases}
\label{tab:peak_diff}
\begin{tabular}{llllll}
\hline
\hline
\multicolumn{2}{l}{\multirow{2}{*}{Object}} & \multicolumn{2}{c}{Full B-scan (soil-surface and object reflections)} & \multicolumn{2}{c}{Object-only B-scan} \\
\multicolumn{2}{l}{} & Before FFDT (dB) & After FFDT (dB) & Before FFDT (dB) & After FFDT (dB) \\
\hline
\multirow{4}{*}{PEC} 
& cylinder (case study in Section~\ref{sec:CaseStudy}) 
& \textcolor{black}{4.67} & \textcolor{black}{$-$45.88} & \textcolor{black}{4.68} & \textcolor{black}{$-$43.31} \\
& rectangular prism 
& \textcolor{black}{5.17} & \textcolor{black}{$-$37.47} & \textcolor{black}{5.16} & \textcolor{black}{$-$37.93} \\
& equilateral triangular prism (0$^\circ$) 
& \textcolor{black}{4.67} & \textcolor{black}{$-$45.89} & \textcolor{black}{4.27} & \textcolor{black}{$-$43.22} \\
& equilateral triangular prism (45$^\circ$) 
& \textcolor{black}{4.73} & \textcolor{black}{$-$31.59} & \textcolor{black}{4.89} & \textcolor{black}{$-$30.43} \\
\hline
\multirow{4}{*}{Concrete} 
& cylinder 
& \textcolor{black}{4.67} & \textcolor{black}{$-$45.91} & \textcolor{black}{4.57} & \textcolor{black}{$-$42.37} \\
& rectangular prism 
& \textcolor{black}{4.67} & \textcolor{black}{$-$45.89} & \textcolor{black}{5.01} & \textcolor{black}{$-$37.64} \\
& equilateral triangular prism (0$^\circ$) 
& \textcolor{black}{4.67} & \textcolor{black}{$-$45.91} & \textcolor{black}{5.07} & \textcolor{black}{$-$37.34} \\
& equilateral triangular prism (45$^\circ$) 
& \textcolor{black}{4.67} & \textcolor{black}{$-$45.88} & \textcolor{black}{4.84} & \textcolor{black}{$-$31.03} \\
\hline
\multirow{2}{*}{Complex object} 
& object 1 
& \textcolor{black}{4.67} & \textcolor{black}{$-$45.86} & \textcolor{black}{4.30} & \textcolor{black}{$-$41.93} \\
& object 2 
& \textcolor{black}{4.67} & \textcolor{black}{$-$45.84} & \textcolor{black}{5.15} & \textcolor{black}{$-$40.85} \\
\hline
\hline
\end{tabular}
\end{table*}

\subsubsection{PEC Objects of Different Geometries} 
In addition to the cylindrical PEC object presented in the case study, FFDT performance is evaluated for a PEC rectangular prism and a PEC equilateral triangular prism at different rotations, as shown in Fig.~\ref{fig:PEC objects comparison}. These objects introduce multiple reflection points that affect the backscattered field, thereby representing more challenging cases for assessing FFDT performance.  Except for the object geometry, all other settings are kept the same as in the case study. 

For the rectangular prism, reflections may originate from the top surface and the four corners, and the dominant reflection depends on the GPR position. To determine the transmitting/receiving angle $\theta$ for applying transfer functions in FFDT at each GPR survey position, $\theta$ is selected to correspond to the dominant reflection associated with the shortest propagation path at that GPR position. Specifically:
\begin{enumerate}
  \renewcommand{\labelenumi}{\alph{enumi})}
  \item If the antenna’s \(x\)-coordinate lies to the left of the object, $\theta$ is set to the angle associated with the top-left corner reflection;
  \item If the antenna’s \(x\)-coordinate lies to the right of the object, $\theta$ is set to the angle associated with the top-right corner reflection;
  \item If the antenna's \(x\)-coordinate lies above the object's top horizontal face,  $\theta = 0^\circ$.
\end{enumerate}
The results before and after FFDT using this $\theta$-selection strategy are shown in Fig.~\ref{fig:PEC objects comparison}(a) and Table~\ref{tab:peak_diff}.  Before FFDT, the maximum Diff between the SimGPR and RealGPR B-scans is \textcolor{black}{5.17} dB. After FFDT, the maximum Diff between transferred B-scan and RealGPR B-scan decreases to \textcolor{black}{$-$37.47}~dB, demonstrating effective domain transfer. Compared with the cylindrical object case, a slightly larger Diff is observed after FFDT. This is because $\theta$ is selected based on the dominant reflection; minor reflections (e.g., edge reflections) may not be transferred using their corresponding transfer functions, which leads to small residual differences between the transferred B-scan and the RealGPR B-scan.  

For the equilateral triangular prism,  a similar $\theta$-selection strategy is adopted in the FFDT process. When a specular reflection path from one face of the prism to the antenna exists, $\theta$ is set to the corresponding angle. Otherwise, $\theta$ is chosen as the angle associated with the prism's vertex closest to the antenna. Results for the prism at 0$^\circ$ and 45$^\circ$ are shown in Figs.~\ref{fig:PEC objects comparison}(b) and \ref{fig:PEC objects comparison}(c), respectively, and are summarized in Table~\ref{tab:peak_diff}.
For the 0$^\circ$ case, the dominant reflections come primarily from the edges, so transfer-function mismatch is limited. The maximum Diff between RealGPR and SimGPR B-scans before and after FFDT are \textcolor{black}{4.67}~dB and \textcolor{black}{$-$45.89}~dB, respectively. When the prism is rotated to 45$^\circ$, dominant reflections may come from both edges and the tilted face. In this case, when $\theta$ is selected based on the tilted face reflection, transfer-function mismatch is introduced for edge reflections, which increases the Diff after FFDT. The maximum Diff before and after FFDT are \textcolor{black}{4.73}~dB and \textcolor{black}{$-$31.59}~dB, respectively. Although the Diff is slightly larger than that in other cases, a reduction of more than 35~dB still demonstrates the effectiveness of FFDT under complex geometries. 

\subsubsection{Dielectric Objects of Different Geometries} 

Besides PEC objects, the performance of FFDT for different dielectric objects is also examined. The same four geometries considered in the PEC object cases are used: a cylinder, a rectangular prism, and an equilateral triangular prism at 0$^\circ$  and 45$^\circ$ rotations. The object dielectric properties are set to those of concrete with 12\% moisture content, which are modeled using a single-pole Debye dispersion with  $\epsilon_{\text{static}}=12.84$, $\epsilon_{\infty}=7.42$, $\tau=0.611~\text{ns}$, and $\sigma_{\mathrm{dc}}=2.06\times10^{-2}~\mathrm{S/m}$~\cite{concrete}. Unlike PEC objects, electromagnetic waves can partially transmit through dielectric objects and scatter from internal boundaries, which introduces more scattering components and increases the complexity of the backscattered field. 

In the FFDT process, the same $\theta$-selection strategy as in the PEC cases is adopted. The maximum Diff between RealGPR B-scans and SimGPR B-scans before and after FFDT are summarized in Table~\ref{tab:peak_diff}. For the full B-scan that includes both soil-surface and object reflections, the Diff between the transferred B-scan and the RealGPR B-scan is reduced to below $-$45~dB, which is smaller than in the PEC cases. This is because the dielectric objects' reflections are weaker than the soil-surface reflection. Consequently, the transferred object reflections, which are more affected by the transfer-function mismatch, have smaller magnitudes, leading to a lower overall Diff. 

To better assess the FFDT performance of the object reflections, object-only cases are also considered. Specifically, Diff is computed between the transferred object-only B-scan and the RealGPR object-only B-scan, with normalization performed using the maximum absolute value across the two B-scans. The results are summarized in Table~\ref{tab:peak_diff}. FFDT effectively reduces the Diff to below $-$30 dB for all PEC and dielectric object-only B-scans, demonstrating its effectiveness in reducing the sim-to-real domain gap.

\subsubsection{\textcolor{black}{Objects with Complex Internal Structures} }

\begin{figure}[tb]
	\centering
    \centerline{\includegraphics[width=1\linewidth]{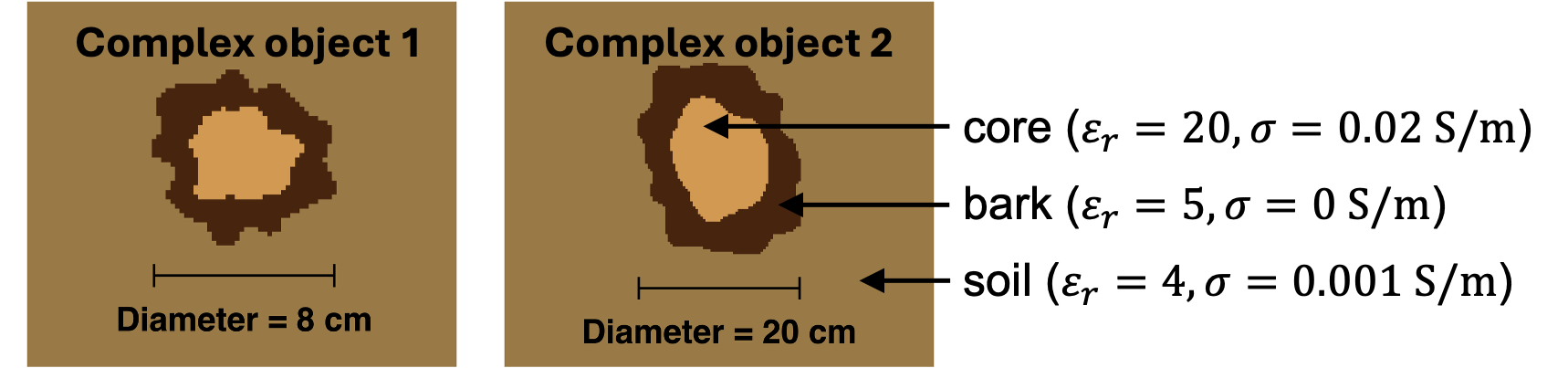}}
    \vspace{-6pt}
	\caption{\textcolor{black}{Cross-sectional view of two root-like objects with complex internal structures.}}
    \vspace{-10pt}
	\label{fig:ComplexObject}
\end{figure}
\textcolor{black}{The performance of FFDT is further evaluated for objects with complex internal structures. Two tree-root-like structures with a high-permittivity core and a lower-permittivity bark layer are modeled, as shown in Fig.~\ref{fig:ComplexObject}. These objects are buried at a depth of 0.2 m in soil. These objects generate more complex scattering than homogeneous dielectric objects.}

\textcolor{black}{The FFDT performance for these two cases is summarized in Table~\ref{tab:peak_diff}. The maximum Diff after FFDT is less than $-$45 dB for the full B-scans and less than $-$40 dB for the object-only B-scans, demonstrating that FFDT remains effective for targets with complex internal structures. This performance is achieved because, although the targets have irregular shapes and internal material contrasts, the dominant reflected components received by the GPR antenna occupy a relatively limited angular range. Therefore, the transfer functions with the estimated $\theta$ within this range can effectively transfer the dominant reflection components from the SimGPR domain to the RealGPR domain.}

\subsection{FFDT Performance with Different Soil Conditions} \label{DiffSoil}
The previous cases considered an object buried in homogeneous soil with a flat surface. In this subsection, the performance of FFDT is evaluated under more realistic subsurface conditions by introducing soil heterogeneity, surface roughness, and layered media. For a controlled analysis, heterogeneous soil with a flat surface and homogeneous soil with a rough surface are investigated separately. In both studies, all other simulation settings are identical to those in the case study described in Section~\ref{sec:CaseStudy}, in which a cylindrical PEC object is buried at a depth of 0.2~m.

\subsubsection{Heterogeneous Soil} \label{HeteroSoilSection}

The heterogeneous soil is generated in gprMax using the  Peplinski mixing model~\cite{Peplinski}. The model parameters are set to a sand fraction of 0.15, a clay fraction of 0.15, a bulk density of 1.1~g/cm$^{3}$, and a sand particle density of 2.3~g/cm$^{3}$. Soil heterogeneity is realized by setting spatially varying soil volumetric water content (VWC). Specifically, the soil is constructed from 50 stochastic soil fractions whose VWC values were sampled within a defined range. The VWC ranges in this study were set to 0\% (homogeneous), 5\%, 10\%, 20\%, and 30\% with the mean VWC fixed at 20\%. For each range, ten random VWC distributions are generated to account for variability in heterogeneous spatial distribution. 

\begin{figure}[tb]
	\centering
    \includegraphics[width=0.9\linewidth]{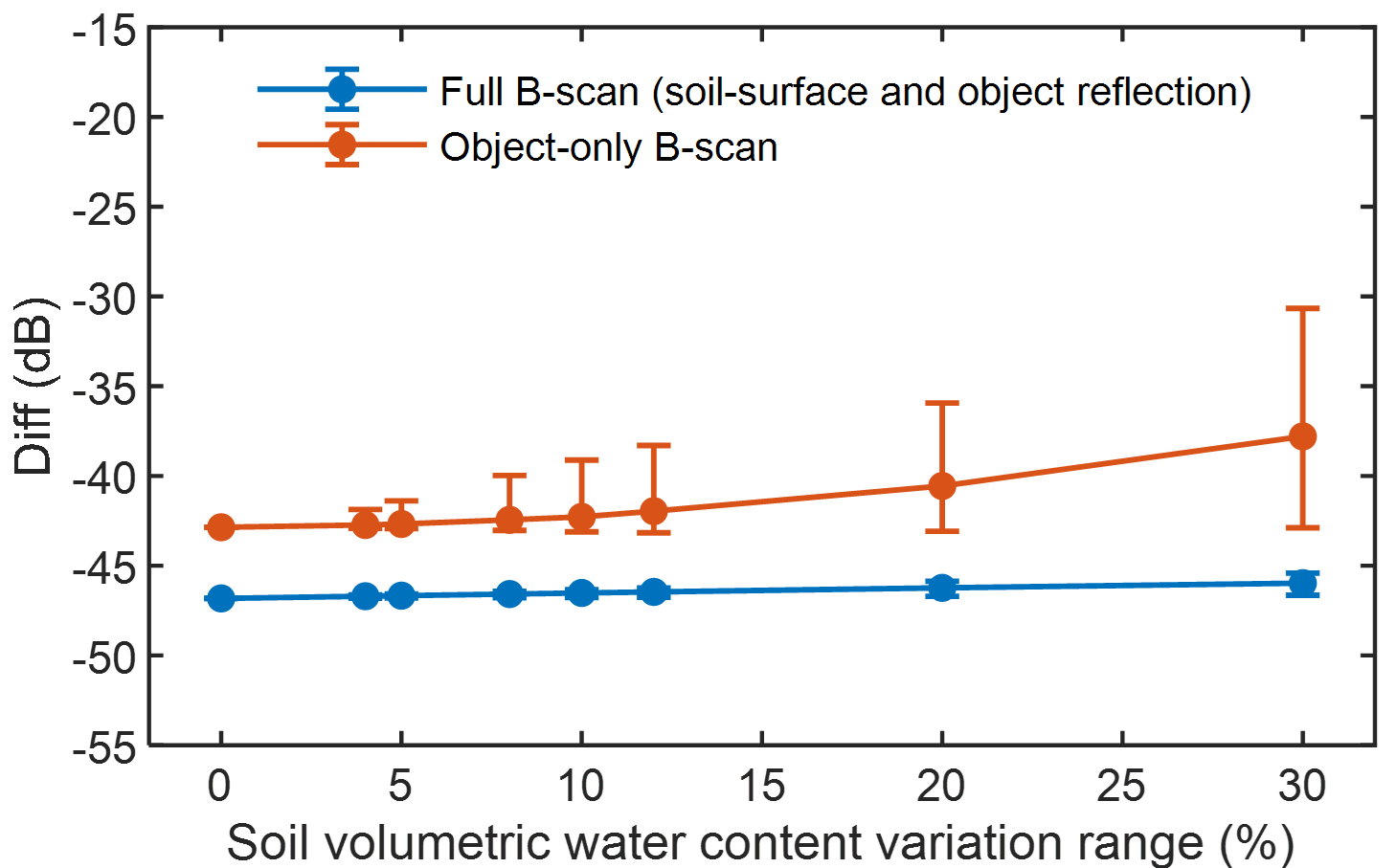}
    \vspace{-6pt}
	\caption{Maximum Diff between between the FFDT-transferred B-scan and the corresponding RealGPR B-scan in heterogeneous soil cases at different soil volumetric water content (VWC) variation ranges. Vertical whiskers indicate the range of maximum Diff across ten heterogeneous spatial distributions for each VWC range.}
    \vspace{-8pt}
	\label{fig:Heterogeneous}
\end{figure}

For FFDT implementation, the soil dielectric property used in ray tracing to estimate $\theta$ is approximated by the average permittivity over the VWC distribution. This approximation is motivated by practical considerations: bulk-averaged soil properties can often be estimated, whereas fine-resolution soil properties are not readily available. For each VWC variation range, the maximum Diff between the FFDT-transferred B-scan and the corresponding RealGPR B-scan is computed across the ten different heterogeneous distributions. These values are then averaged to obtain the final Diff metric, which are shown in Fig.~\ref{fig:Heterogeneous}. The vertical whiskers in the figure indicate the range of maximum Diff across the ten heterogeneous distributions.  The Diff increases as the VWC variation range increases. This trend is expected because stronger heterogeneity causes the local permittivity to deviate more from the averaged permittivity, which increases errors in $\theta$ estimation and, consequently, the transfer-function mismatch. This effect is more obvious in the object-only B-scan than in the full B-scan, because the transfer of the object reflection is more sensitive to $\theta$ errors, whereas the soil-surface reflection is always associated with $\theta_s=0^\circ$. Nevertheless, even for the most challenging case with 30\% VWC variation range, the FFDT-transferred B-scans still closely match the RealGPR B-scans, with differences below $-$30~dB. These results demonstrate the effectiveness of FFDT in heterogeneous soil environments.

\subsubsection{Soil Surface Roughness} 

Homogeneous soil with different surface roughness is constructed using the built-in roughness function in gprMax. The soil parameters are the same as those in  Section~\ref{HeteroSoilSection}, but the VWC is fixed at 20\% to model the homogeneous medium. Surface roughness is quantified by the roughness height, defined as the difference between the maximum and minimum surface elevations. In this analysis, the roughness height varies from 0 (flat) to 10~cm in 2~cm increments. To increase the diversity of the roughness profiles, ten random surface realizations are generated for each roughness height. 

\begin{figure}[tb]
	\begin{center}
		\begin{tabular}{c@{ }}
			\includegraphics[width=0.9\linewidth]{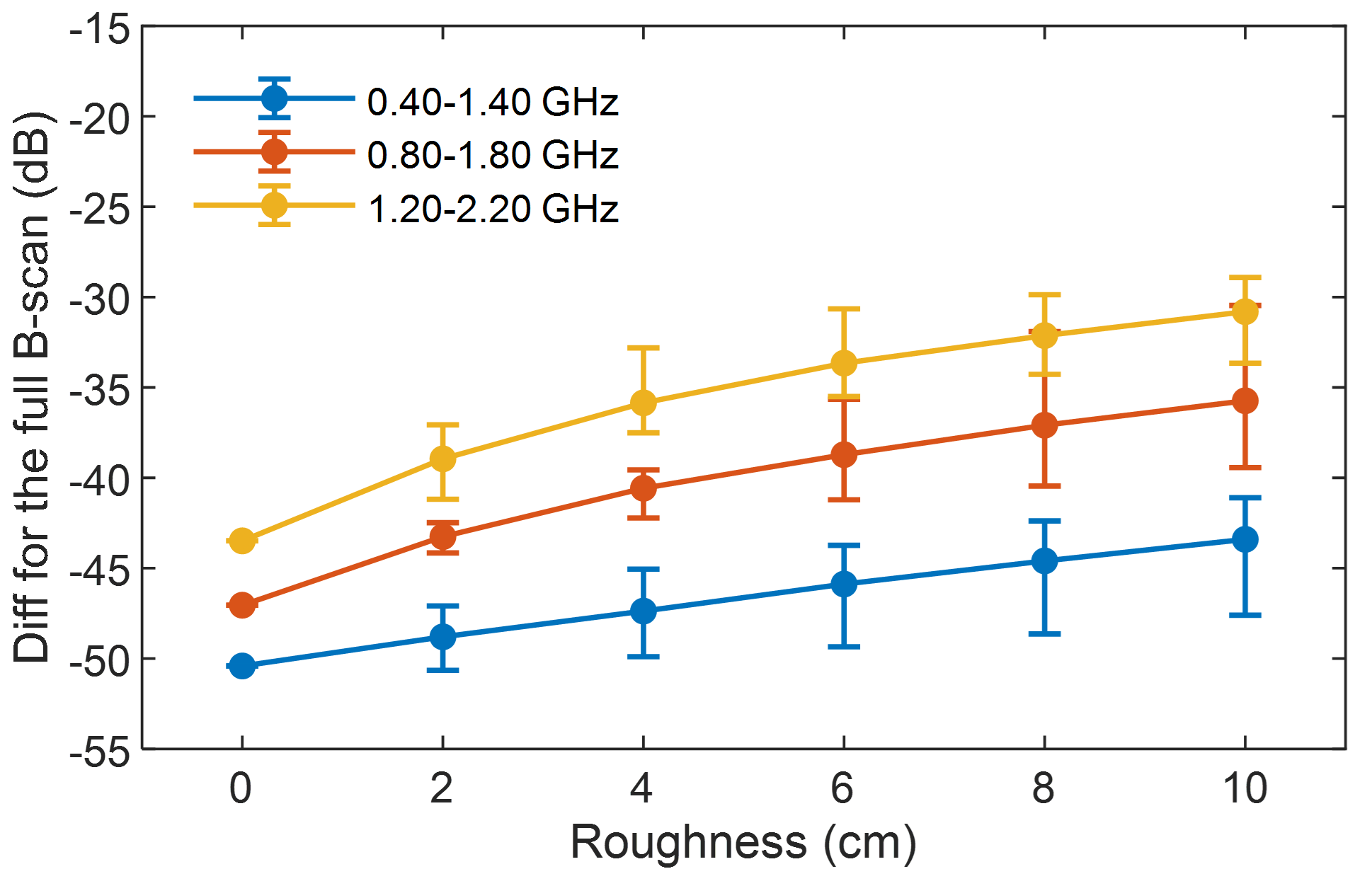}
			\footnotesize{(a)} \\
			\includegraphics[width=0.9\linewidth]{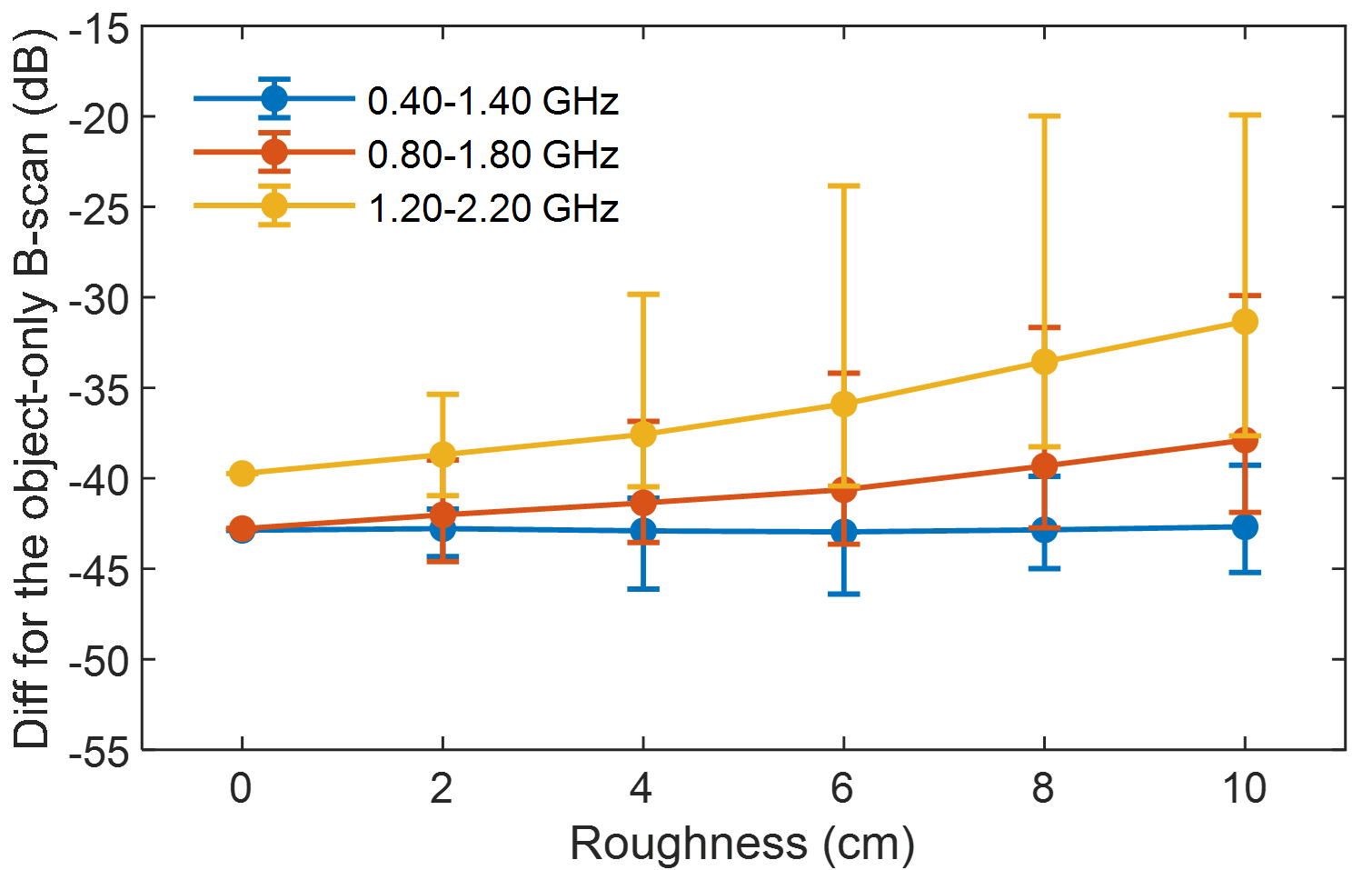}
			\footnotesize{(b)}  \\
		\end{tabular}
	\end{center}
	\caption{Maximum Diff between the FFDT-transferred B-scan and the corresponding RealGPR B-scan for soil with different surface roughness heights. Vertical whiskers indicate the range of maximum Diff across ten roughness profiles for each roughness height. (a) Full B-scan. (b) Object-only B-scan.}
    \vspace{-4pt}

	\label{fig:Roughness}
\end{figure}

During the FFDT implementation, the soil surface is approximated as flat to determine $\theta$. This approximation is motivated by practical consideration: in field deployments, the true surface profile is typically unknown and difficult to measure accurately, and a locally flat surface approximation is often adopted. For each roughness height, the maximum Diff between the FFDT-transferred B-scan and the corresponding RealGPR B-scan is computed across the ten different roughness realizations, and these values are averaged to obtain the final Diff metric. The maximum Diff for the full B-scan and the object-only B-scan is shown in Figs.~\ref{fig:Roughness}(a) and ~\ref{fig:Roughness}(b), respectively, where the vertical whiskers indicate the range of maximum Diff across the ten roughness realizations. As roughness sensitivity is frequency dependent, the results are analyzed for three different frequency bands with the same bandwidth: 0.4-1.4~GHz, 0.8-1.8~GHz, and 1.2-2.2~GHz.

As shown in Fig.~\ref{fig:Roughness}, for each frequency band, Diff increases with roughness height, which indicates degraded FFDT performance. The trend is expected: when the roughness height becomes comparable to the wavelength, the flat-surface approximation introduces larger ray-path and $\theta$ estimation errors, which increases transfer-function mismatch.  The degradation is less pronounced at lower frequencies (e.g., 0.4-1.4~GHz), where the longer wavelength reduces sensitivity to roughness. These observations are consistent with the classical Rayleigh roughness criterion that a surface may be treated as effectively smooth when the height deviation from a plane is less than $\frac{\lambda}{8\cos\theta}$ ~\cite{rayleigh}. Overall, FFDT reduces Diff to below $-$30~dB in most cases, demonstrating that FFDT remains effective even in the presence of surface roughness.

\begin{figure}[tb]
	\centering
    \includegraphics[width=1\linewidth]{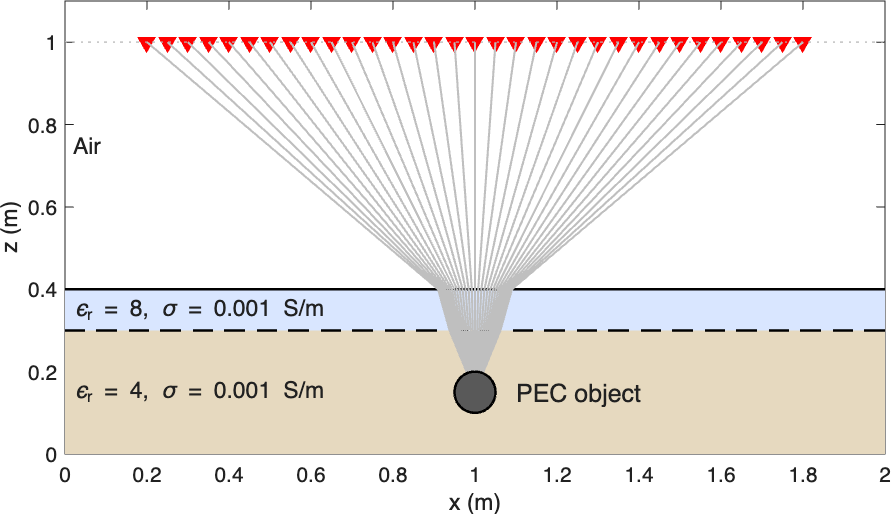}
    \vspace{-8pt}
	\caption{\textcolor{black}{Two-layer medium scenario used to evaluate FFDT performance in layered media.}}
    \vspace{-8pt}
	\label{fig:Layer_Medium}
\end{figure}

\subsubsection{\textcolor{black}{Layered medium}}
\textcolor{black}{To evaluate the applicability of FFDT beyond single-layer medium cases, a two-layer medium scenario is modeled, as shown in Fig.~\ref{fig:Layer_Medium}. For the FFDT implementation, the transmitting/receiving angle $\theta$ is estimated using ray tracing with Snell’s law across the air–medium and interlayer interfaces. Before FFDT, the maximum Diff between the SimGPR and RealGPR B-scans is $-$2.11 dB for the full B-scan and $-$2.25 dB for the object-only B-scan. After applying FFDT, the maximum Diff is effectively reduced to $-$42.71 dB for the full B-scan and $-$43.06 dB for the object-only B-scan. These values are comparable to those obtained in the single-layer case study. Layered media with different layer depths and material properties are also examined, and similar results are obtained. These results demonstrate that FFDT maintains high performance in layered media as long as the transmitting/receiving angle is well estimated.}

\subsection{Determination of the Far-Field Criterion} \label{FarFieldCondition}

\begin{figure}
	\begin{center}
			\includegraphics[width=0.95\linewidth]{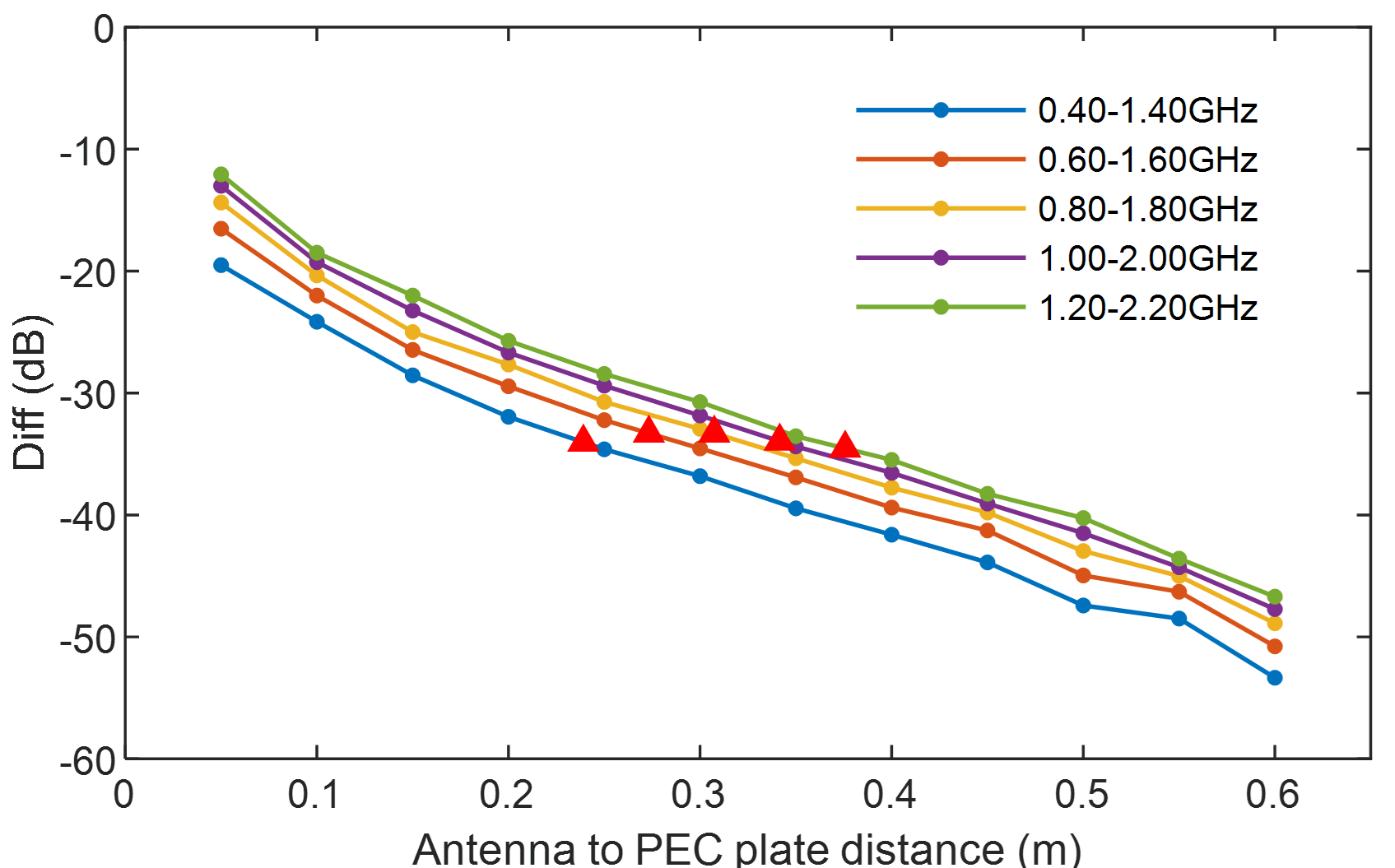}
            \vspace{-4pt}
	\end{center}
	\caption{FFDT performance at different antenna-to-PEC plate distances, quantified by the maximum Diff between the FFDT-transferred A-scan and the RealGPR A-scan.}
    \vspace{-4pt}
	\label{fig:FF comparison}
\end{figure}
To determine the far-field criterion under which FFDT can be reliably applied, FFDT performance is evaluated at different antenna-to-PEC plate distances across multiple frequency bands.  Specifically, using the SimGPR and RealGPR antenna configurations described in Section~\ref{sec:CaseStudy}, A-scans are acquired with the antenna centered above a large PEC plate at distance ranging from 0.05~m to 0.6~m in 0.05~m increments. For each distance, SimGPR data were transferred using FFDT over five frequency bands with the same bandwidth: 0.4-1.4 GHz, 0.6-1.6~GHz, 0.8-1.8~GHz, 1.0-2.0~GHz, and 1.2-2.2~GHz. The maximum Diff between the normalized transferred A-scan and the RealGPR A-scan in the same band is computed. The results are shown in Fig.~\ref{fig:FF comparison}. At a fixed distance, Diff increases with frequency, representing degraded FFDT performance. At a fixed frequency band, Diff decreases with distance, showing improved FFDT performance. 

This trend is consistent with the Fraunhofer far-field criterion, given by
$R \ge \frac{2D^2}{\lambda}$ \cite{AntennaBook},
where $D$ is the maximum antenna dimension and $\lambda$ is the wavelength. In the present setup, the RealGPR system comprises two antenna boxes with an overall maximum dimension of approximately 0.16~m. The Fraunhofer distance computed at the highest frequency for each frequency band is marked by red triangles in Fig.~\ref{fig:FF comparison}. When the antenna-to-PEC distance exceeds this distance, Diff remains consistently below $-$30~dB. Therefore, we adopt
$R \ge \frac{2D^2}{\lambda_h}$
as the far-field criterion for reliable FFDT operation, where $\lambda_h$ denotes the wavelength at the highest frequency of the operating band.

\begin{figure}[tb]
	\begin{center}
		\includegraphics[width=0.95\linewidth]{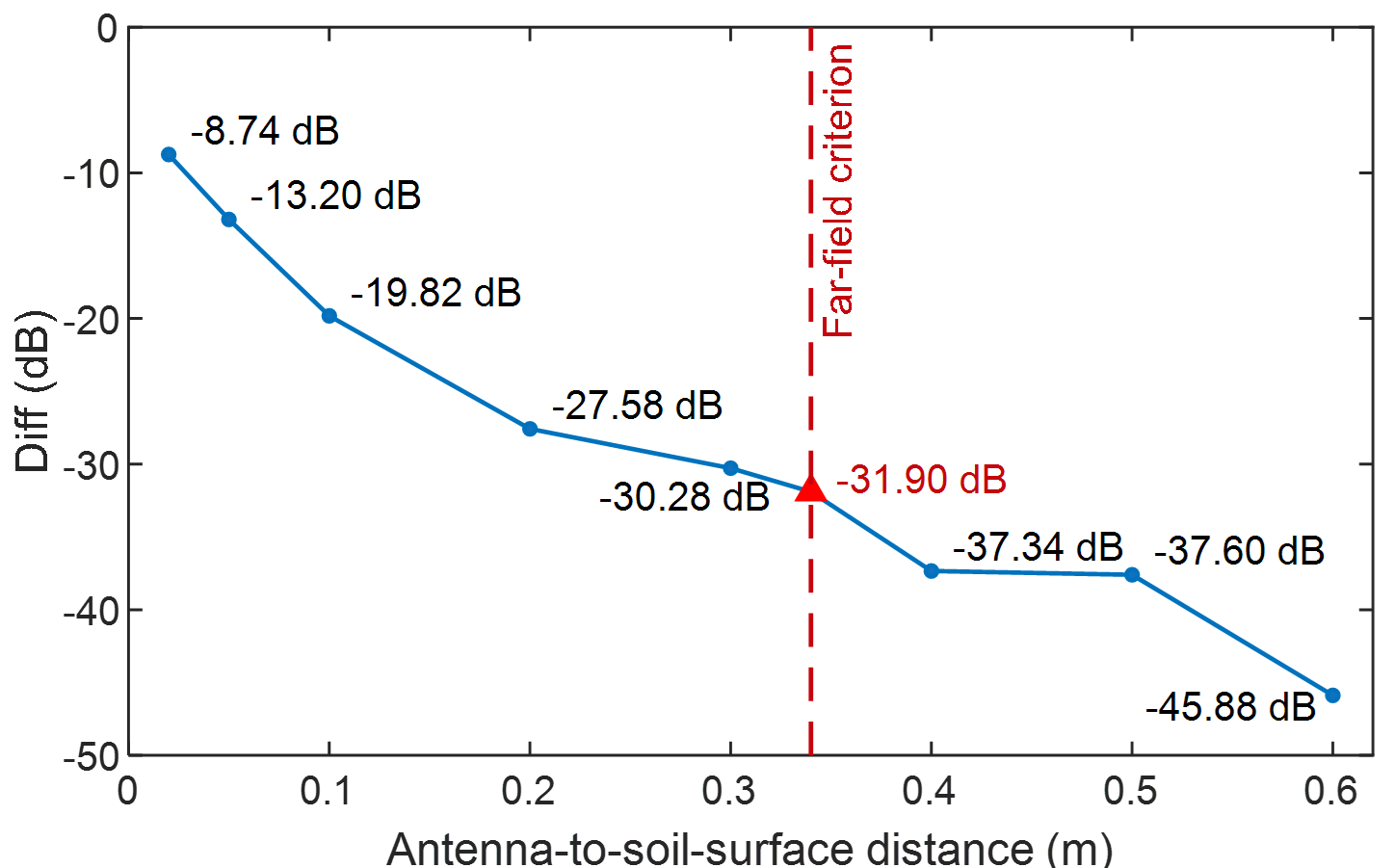}
        \vspace{-8pt}
	\end{center}
	\caption{FFDT performance at different antenna-to-soil-surface distances in the case-study scenario in Section~\ref{sec:CaseStudy}, quantified by the maximum Diff between the FFDT-transferred B-scan and the RealGPR B-scan.}
    \vspace{-8pt}
	\label{fig:revision near-field limitation}
\end{figure}

\textcolor{black}{To further evaluate the performance of FFDT when the far-field condition is not strictly satisfied, we conduct a comparative study based on the case-study configuration in Section~\ref{sec:CaseStudy}, but vary the antenna-to-soil surface distances from 0.02, 0.05, 0.1, 0.2, 0.3, 0.4, 0.5 and 0.6 m while keeping all other parameters unchanged. The performance at the far-field threshold of 0.34 m is also examined and used as a reference. The maximum Diff values between the transferred B-scan and the RealGPR B-scan in these cases are summarized in Fig.~\ref{fig:revision near-field limitation}. As expected, the maximum Diff increases as the antenna-to-surface height decreases. Same as observed in the antenna-to-PEC plate cases, when the far-field criterion is satisfied, the maximum Diff after FFDT remains below $-$30 dB, representing good agreement between the FFDT-transferred B-scan and the RealGPR B-scan. When the criterion is not satisfied, the Diff increases to above $-$30 dB, reflecting the degradation caused by near-field effects that are not captured by the current far-field transfer-function model. Interestingly, even in the near-field case with an antenna-to-surface distance of 0.10 m, the Diff remains around $-$20 dB. This suggests that although FFDT is less accurate in near-field scenarios, the transferred data may still retain sufficient domain alignment for basic deep-learning tasks, such as target detection and localization, where exact waveform matching is less critical than preserving the main reflection signature. }

\begin{figure}[tb]
	\begin{center}
		\includegraphics[width=1\linewidth]{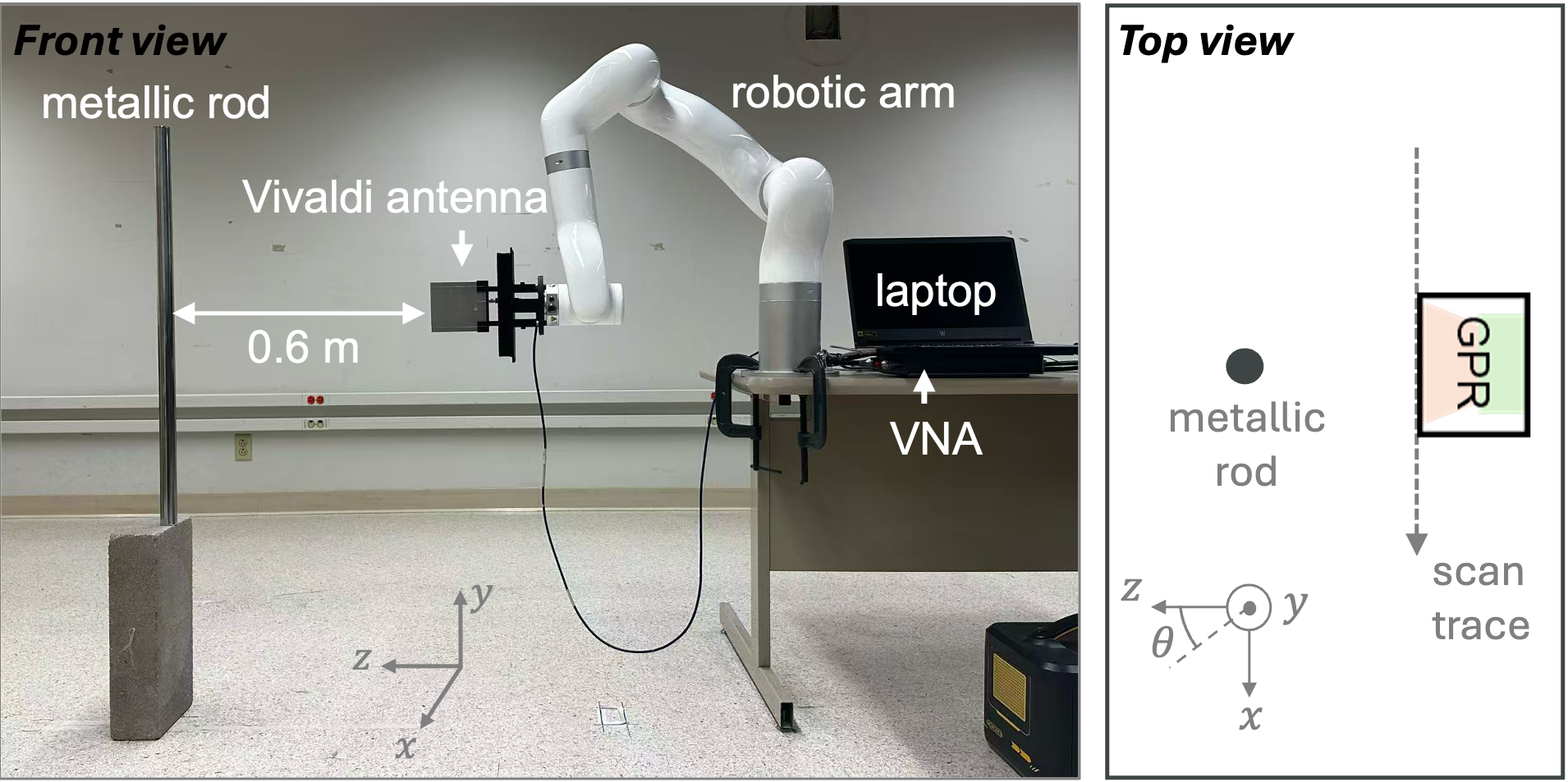}
        \vspace{-8pt}
	\end{center}
	\caption{Experimental setup for B-scan acquisition of a metallic rod in air using the custom VNA+Vivaldi GPR setup. }
    \vspace{-8pt}
	\label{fig:experiment setup example}
\end{figure}

\begin{figure*}[tb]
	\begin{center}
			\includegraphics[width=1\linewidth]{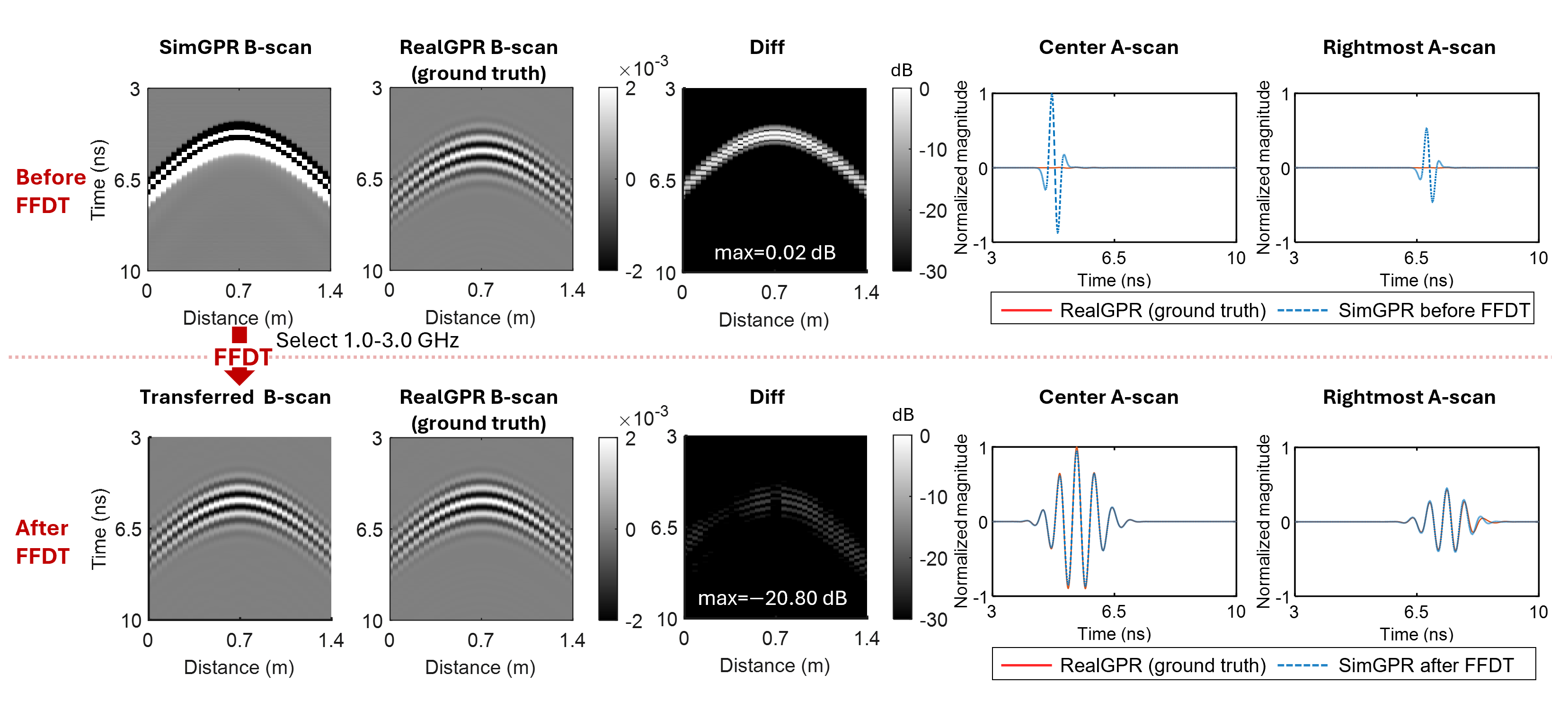}
            \vspace{-20pt}
	\end{center}
	\caption{Comparison of the SimGPR B-scan, FFDT-transferred B-scan, and RealGPR B-scan for the scenario in Fig. \ref{fig:experiment setup example}. Time gating is applied to retain only the object reflection for comparison. }
    \vspace{-0pt}
	\label{fig:sim2realDIFF}
\end{figure*}

\textcolor{black}{The degradation in the near-field region is mainly due to the far-field point-source approximation used in the GPR system model described in Section~\ref{GPRmodel}. In this model, the antenna is represented as an equivalent point radiator/receiver with an angle-dependent radiation pattern. This approximation is valid when the sensing region is sufficiently far from the antenna, i.e., in the Fraunhofer far-field region, where the angular field distribution is essentially independent of the antenna-to-target distance. In contrast, when the sensing region lies in the radiating near-field region, the field distribution depends on both the radial distance and the angular direction from the antenna. In this case, spatial current/field distribution over the finite antenna aperture must be considered to model both distance- and angle-dependent radiation behavior \cite{AntennaBook}. Because these distance-dependent antenna radiation effects are not included in the current far-field point-source formulation, degradation in the near-field region is expected.}

\section{Measurement Results}\label{sec4}
Three experimental studies are presented to validate FFDT performance. First, in Subsections~\ref{MeasurementDemo1} and~\ref{GP8800}, the FFDT-transferred B-scan is directly compared with RealGPR measurements acquired in the same scenario using two different GPR systems. Second, because FFDT is primarily intended to generate RealGPR-domain datasets from simulation for training DL models, in Subsection~\ref{MeasurementDemo2}, FFDT is applied to construct a sim-to-real transferred dataset for training a DL model for clutter removal. The model's performance is evaluated on RealGPR B-scans to demonstrate the effectiveness of FFDT for RealGPR-domain dataset generation. 

\subsection{FFDT Validation Using a Custom GPR System} \label{MeasurementDemo1}

In the scenario shown in Fig.~\ref{fig:experiment setup example}, B-scans are simulated using a SimGPR in gprMax and measured using a RealGPR system. A metallic rod with a diameter of 0.032~m and a length of 0.79~m is placed in air at a standoff distance of 0.6~m from the GPR antenna. A B-scan is collected over a scan trace of 1.4~m with a step size of 0.04~m. This experimental validation is conducted in air because soil properties cannot be replicated exactly between simulation and measurement, whereas an air setup provides a well-controlled environment for evaluating \textcolor{black}{the FFDT mechanism by examining whether the calibrated SimGPR and RealGPR transfer functions can accurately map radar signals from one system domain to the other in the same environment.}

SimGPR is modeled in gprMax using closely spaced Hertzian dipole and field probe as TX and RX, respectively. A Ricker waveform with a 2~GHz center frequency is used for excitation. The transfer functions of the SimGPR are derived following the procedure described in Section~\ref{sec:CaseStudy}. 

RealGPR is implemented as a stepped-frequency radar by connecting a Vivaldi antenna backed by RF absorber to a Copper Mountain Technologies M5090 vector network analyzer (VNA) controlled by a laptop. The VNA collects 201 frequency points from 1~GHz to 3~GHz with 0~dBm output power and a 1~kHz IF bandwidth. During B-scan acquisition, the antenna is mounted on a robotic arm to ensure precise positioning. The transfer functions of the RealGPR are derived following a procedure similar to that in Section~\ref{sec:CaseStudy}. A 1.83~m $\times$ 1.83~m copper plate is used as the calibration target. Using the robotic arm, GPR measurements are acquired at antenna-to-plate distances from 0.66~m to 0.88~m in 0.02~m increments; for each distance, $\theta$ is swept from $-50^{\circ}$ to $+50^{\circ}$ with a step size of $5^{\circ}$. \textcolor{black}{Since the measurements are conducted in an indoor environment where reflections from nearby objects lead to environmental clutter, we adopt three steps to reduce the influence of environmental clutter and improve the robustness of the calibration process:}
\begin{enumerate}
  \renewcommand{\labelenumi}{\alph{enumi})}
  \item \textcolor{black}{A partially controlled indoor environment is used for the calibration measurements. The PEC plate and antenna are placed in an open area, with the nearest surrounding objects, such as tables, chairs, and walls, located approximately 1.5 m away. This distance is about twice the average antenna-to-plate calibration distance. Therefore, the dominant clutter reflections from surrounding objects arrive much later in time than the PEC-plate reflection, allowing the PEC-plate reflection to be sufficiently separated in the time domain.}
  \item \textcolor{black}{Background subtraction and time gating are applied to isolate the PEC-plate reflection. Specifically, we acquire both the PEC-plate measurement and the background measurement after removing the PEC plate. The background response is first subtracted from the PEC-plate measurement to suppress the dominant environmental clutter and antenna coupling. Then, a time-domain gate is applied to extract the PEC-plate reflection for subsequent calibration of the system transfer functions. Through these two processing steps, the environmental clutter in the PEC-plate calibration signal is substantially reduced.}
  \item \textcolor{black}{The background-subtracted, time-gated calibration signals acquired at the 12 antenna-to-plate distances are used to form an overdetermined system of equations, as shown in Eq.~\eqref{eq:matrix_calibration}, for solving the system transfer functions at each $\theta$. This multi-distance formulation further reduces the influence of residual clutter in any individual PEC-plate measurement on the final estimated system transfer functions.}
\end{enumerate}
The RealGPR and SimGPR transfer functions are subsequently used for FFDT implementation.

The SimGPR B-scan, FFDT-transferred B-scan, and RealGPR B-scan from 1~GHz to 3~GHz are compared in Fig.~\ref{fig:sim2realDIFF}. Before FFDT, the maximum Diff between the normalized SimGPR and RealGPR B-scans is \textcolor{black}{0.02}~dB; after FFDT, it decreases to $-\textcolor{black}{20.80}$~dB. The remaining differences are primarily due to small positioning errors during transfer function calibration and B-scan acquisition. Overall, the reduction in Diff of \textcolor{black}{20.82}~dB confirms that FFDT effectively reduces the sim-to-real domain gap.   

\begin{figure}[tb]
	\begin{center}
			\includegraphics[width=1\linewidth]{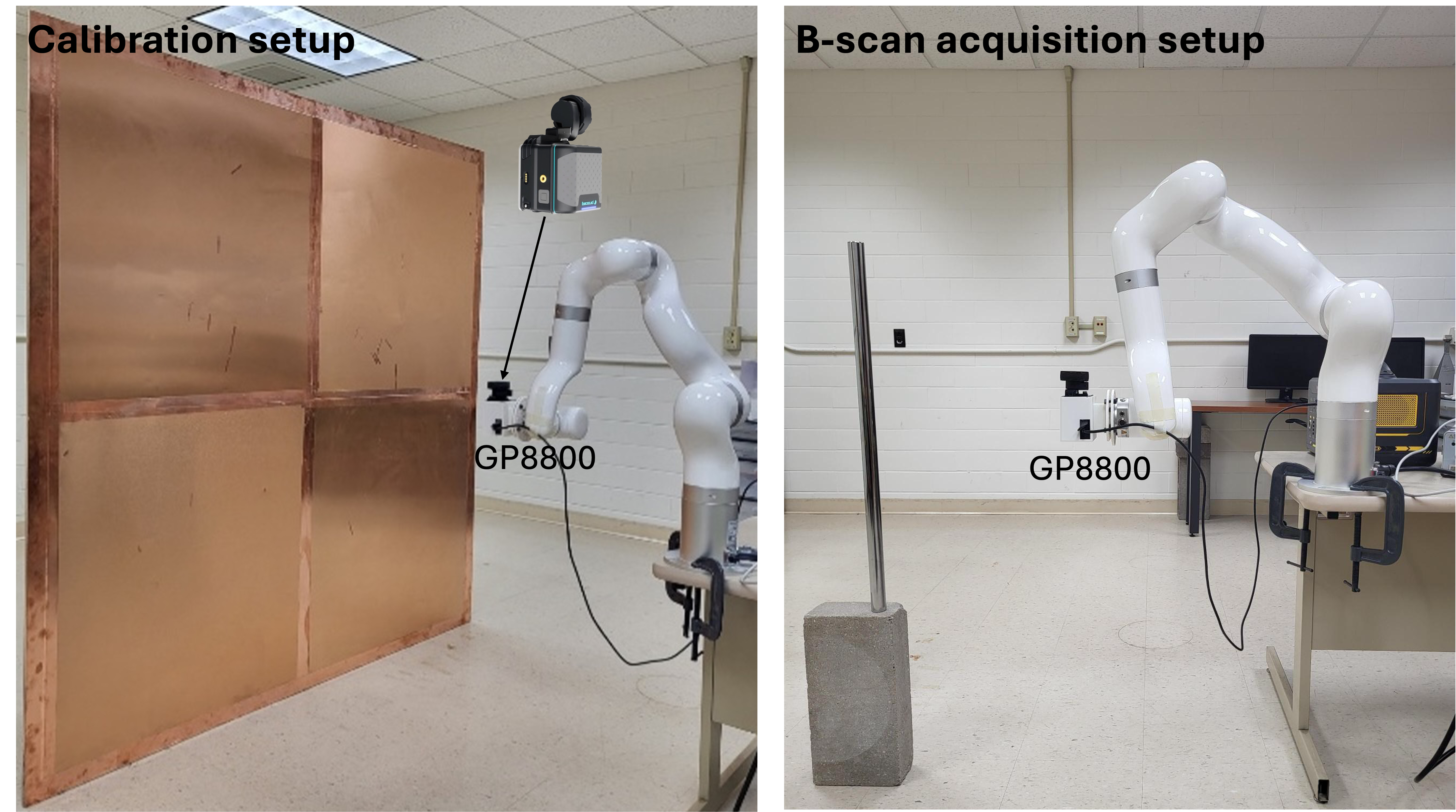}
            \vspace{-8pt}
	\end{center}
	\caption{Experimental setup for calibration and B-scan data acquisition using the Proceq GP8800~\cite{ScreeningEagleGP8800}. }
                \vspace{-8pt}
	\label{fig:GP8800_Exp}
\end{figure}

\begin{figure*}[tb]
	\begin{center}
			\includegraphics[width=1\linewidth]{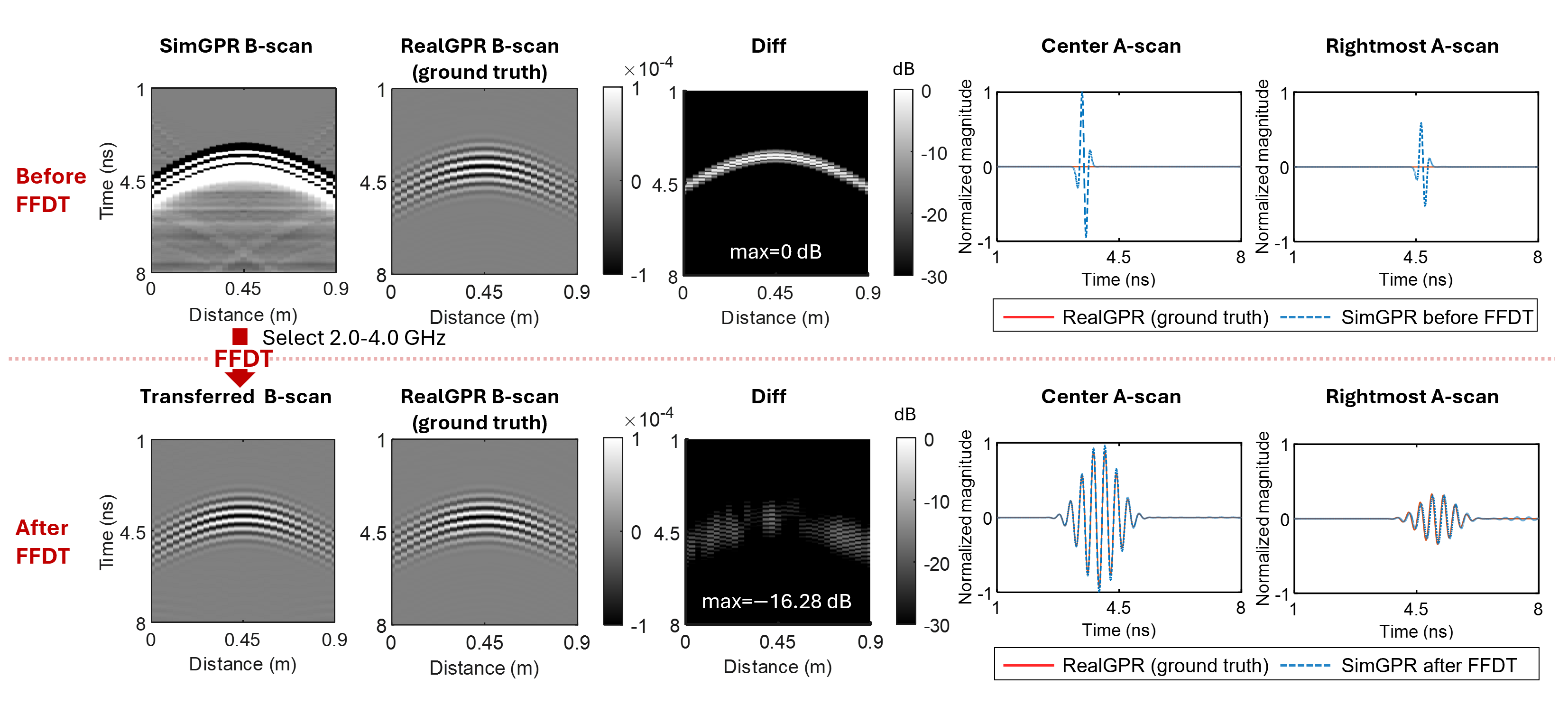}
            \vspace{-16pt}
	\end{center}
	\caption{Comparison of the SimGPR B-scan, FFDT-transferred B-scan, and RealGPR (GP8800) B-scan for the scenario in Fig. \ref{fig:GP8800_Exp}. Time gating is applied to retain only the object reflection for comparison.}
    \vspace{-0pt}
	\label{fig:GP8800_Results}
\end{figure*}

\subsection{\textcolor{black}{Experimental Validation Using a Commercial GPR System}} \label{GP8800}
\textcolor{black}{To further evaluate FFDT using practical GPR hardware, an additional experimental study is conducted using an off-the-shelf GPR system, the Proceq GP8800 \cite{ScreeningEagleGP8800}. The FCC compliance test report specifies a $-$10-dB radiated-emission frequency range of 1.92–5.96 GHz and describes dipole antennas as the TX and RX antennas \cite{ACL2019GP8800}. The TX and RX antennas have a separation of 3.6 cm \cite{Solla2024Rebar}. Because this separation is small relative to the sensing distance considered in this experiment, the GP8800 is treated as a quasi-monostatic system for FFDT validation.}

\textcolor{black}{Since FFDT relies on calibrated system transfer functions, repeatability of the RealGPR response is important. Repeated GP8800 measurements acquired under identical conditions show a maximum Diff of $-$13.69 dB, indicating noticeable measurement-to-measurement magnitude and phase variations. To reduce this variation, an antenna-coupling-based correction is applied to all GP8800 measurements. Specifically, the antenna-coupling response of each measurement is compared with that of a fixed reference measurement to determine a magnitude scaling factor and phase/time-shift correction, which are then applied to the complete measured signal. After correction, the maximum Diff between repeated measurements decreases to $-$33.91 dB, demonstrating substantially improved measurement consistency.}

\textcolor{black}{The SimGPR is implemented in gprMax using a Hertzian dipole and a field probe as the TX and RX, respectively, with a separation of 3.6 cm. A Ricker waveform with a center frequency of 3 GHz is used for excitation. The transfer functions of both the SimGPR and GP8800 are calibrated using an electrically large PEC plate described in Section~\ref{MeasurementDemo1}, as shown in Fig.~\ref{fig:GP8800_Exp}. Calibration measurements are obtained at antenna-to-plate distances from 0.44 to 0.60 m in 0.02 m increments and angles from $-$45$^\circ$ to 45$^\circ$ in 5$^\circ$ increments. The transfer functions are calibrated over the common 2-4 GHz frequency range and subsequently used for FFDT implementation.}

\textcolor{black}{For FFDT validation, the experimental setup is shown in Fig.~\ref{fig:GP8800_Exp}. The same metallic rod as in Section~\ref{MeasurementDemo1} is positioned in air at a standoff distance of 0.45 m. SimGPR and RealGPR (GP8800) B-scans are obtained over a scan trace of 0.90 m with a spatial step of 0.03 m. The SimGPR B-scan, FFDT-transferred B-scan, and measured GP8800 B-scan over 2–4 GHz are compared in Fig.~\ref{fig:GP8800_Results}. Before FFDT, the maximum Diff between the normalized SimGPR and GP8800 B-scans is 0 dB. After FFDT, the maximum Diff decreases to $-$16.28 dB. The transferred B-scan closely reproduces the reflection characteristics of the measured GP8800 response, demonstrating that FFDT substantially reduces the sim-to-real domain gap for this commercial GPR system.}

\textcolor{black}{The remaining difference is slightly larger than that obtained using the custom VNA+Vivaldi system in Section~\ref{MeasurementDemo1}, which is due lower repeatability of the GP8800. Even after antenna-coupling-based correction, repeated GP8800 measurements show a maximum Diff of $-$33.91 dB, which is higher than $-$51.06 dB for our custom system. Nevertheless, the reduction from 0 to $-$16.28 dB provides experimental evidence that FFDT can be applied to practical GPR hardware.}

\subsection{FFDT-Enabled GPR Clutter-Removal DL Model} \label{MeasurementDemo2}

To demonstrate that FFDT can leverage simulated data to construct datasets that better match the RealGPR domain, and thereby improving the performance of DL models  on RealGPR measurements, FFDT is applied to a classical task: clutter removal in GPR radargrams. Specifically, the same DL model is trained on (a) SimGPR datasets and (b) FFDT-transferred datasets, and its performance is compared on RealGPR B-scans. The SimGPR and RealGPR systems are the same as those implemented in Section~\ref{MeasurementDemo1}.

\textbf{Dataset generation.} A dataset is generated in gprMax using SimGPR. A metallic rod is buried in a homogeneous medium representing sandy soil with a relative permittivity of 3 and a conductivity of 0.001~S/m. For each scenario, the rod's burial depth is randomly selected from 0.03~m to 0.16~m, and the horizontal position along the $x$-axis is randomly selected from $-0.4$~m to $+0.4$~m relative to the center of the scan range. The soil surface roughness height varies from 1~cm to 16~cm, and the antenna-to-surface distance varies from 0.1~m to 0.8~m. B-scans are acquired over a 1.4~m scan length with a 4~cm step size.   For each scenario, a clutter-only B-scan is also simulated by maintaining the same soil and surface conditions but removing the buried object. A clutter-free B-scan is then obtained by subtracting the clutter-only B-scan from the raw B-scan. The raw B-scan and the corresponding clutter-free B-scan form a data pair for supervised learning. All B-scans exclude $H_a$. A total of 1,300 data pairs are generated and then doubled to 2,600 pairs through horizontal-flip augmentation. 

Two versions of datasets are constructed. The dataset obtained from simulation is referred to as the SimGPR dataset, whereas the dataset after applying FFDT is referred to as the FFDT-transferred dataset. Each B-scan is resized to 64$\times$2048 and normalized to $[-1,1]$ using the maximum value of the raw B-scan in the data pair.

\begin{figure}[tb]
	\begin{center}
			\includegraphics[width=1\linewidth]{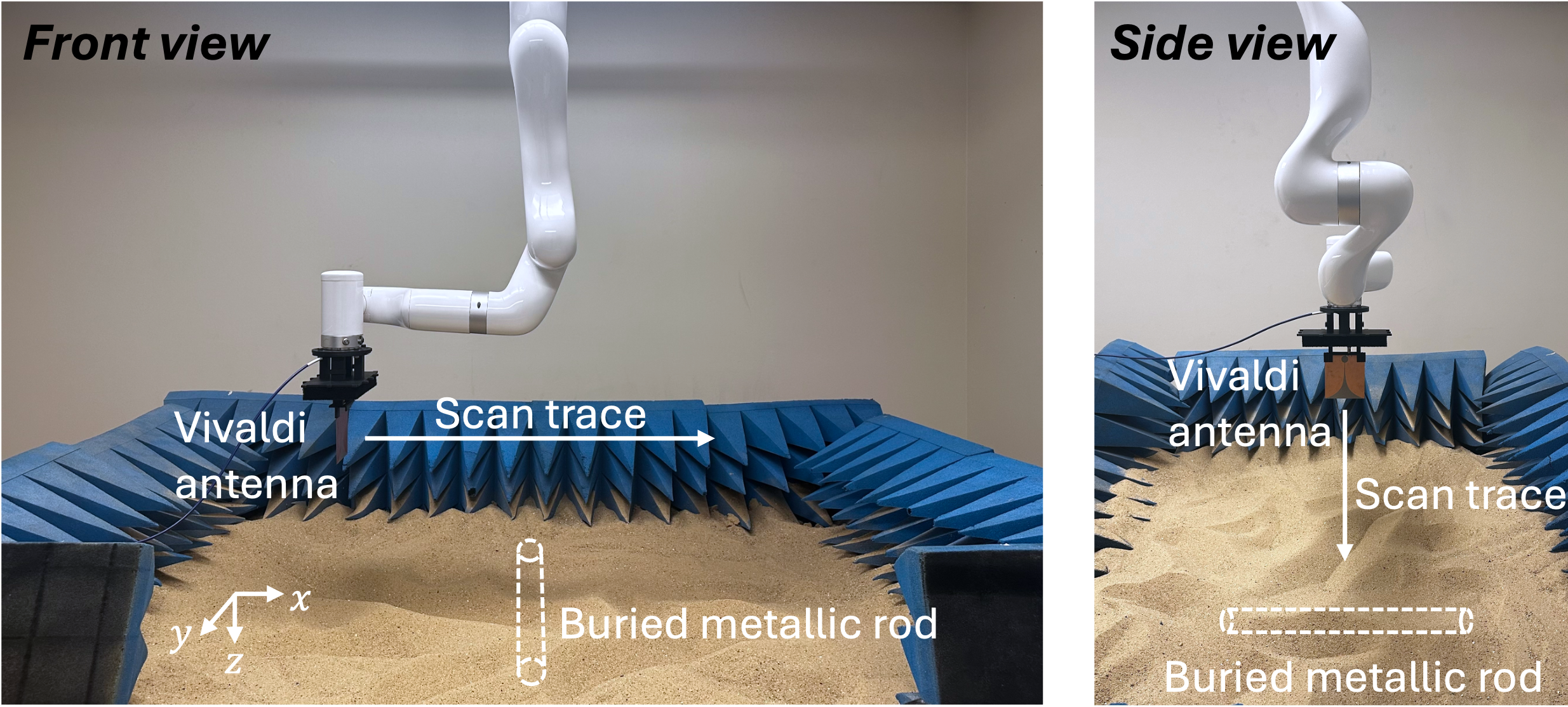}
            \vspace{-8pt}
	\end{center}
	\caption{Experimental setup for B-scan acquisition of a metallic rod buried in a sand testbed using the RealGPR. }
                \vspace{-0pt}
	\label{fig:sandboxBscan}
\end{figure}

\begin{figure}
    \includegraphics[width=0.97\linewidth]{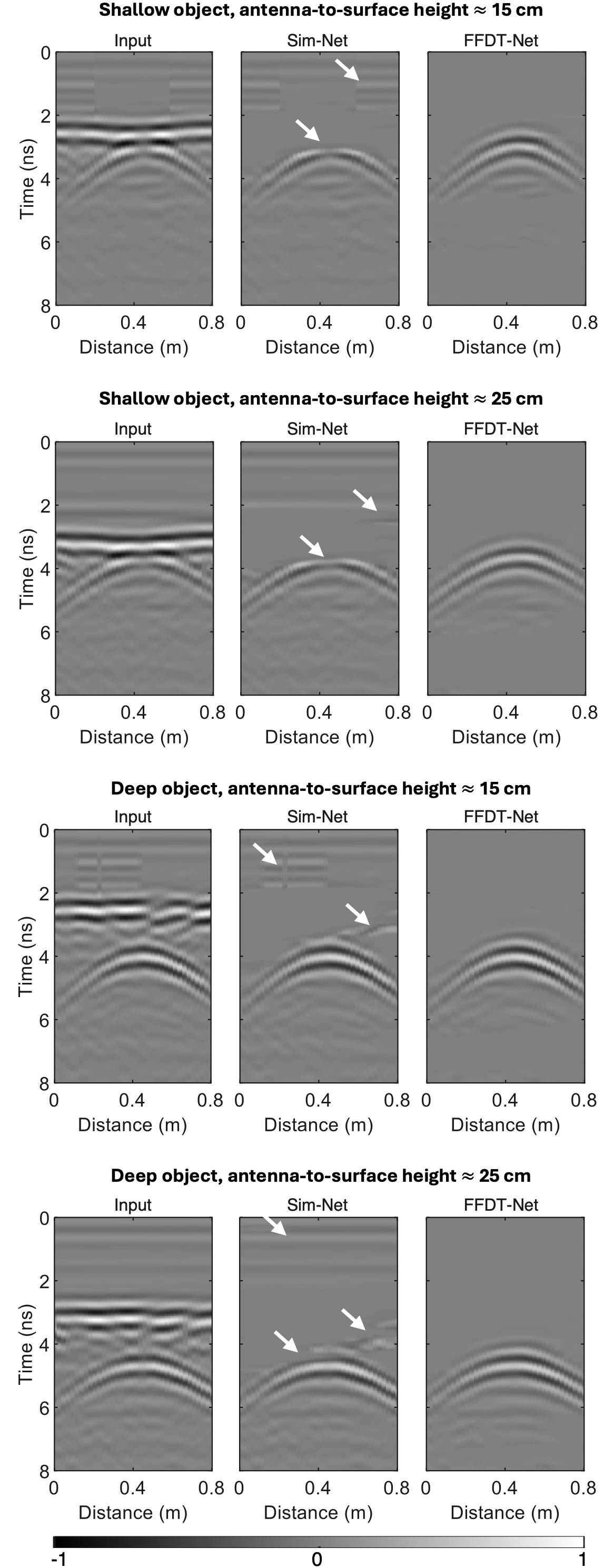}
            \vspace{-13pt}
	\caption{Comparison of clutter-removal performance of the Sim-Net and FFDT-Net on RealGPR B-scans. The Sim-Net and the FFDT-Net are trained on the SimGPR dataset and the FFDT-transferred dataset, respectively. Compared to the Sim-Net, the FFDT-Net shows substantially improved performance in removing clutter and restoring object reflections, which is attributed to improved training domain alignment enabled by FFDT.}            
    \vspace{-15pt}
	\label{fig:network results}
\end{figure}

\textbf{DL model training.} The classic U-Net \cite{unet} is adopted as the baseline model for the clutter removal task. U-Net is a widely used encoder-decoder architecture for image-to-image learning and is commonly used as a baseline for fair comparisons across datasets. The U-Net is separately trained on the SimGPR dataset and the FFDT-transferred dataset to learn a mapping from the input raw B-scan to the output clutter-free B-scan. The U-Net model trained on the SimGPR dataset is referred to as the Sim-Net, while the one trained on the FFDT-transferred dataset is referred to as the FFDT-Net.

Following~\cite{clutter1}, the loss function used in the training process is defined as the combination of mean absolute error (MAE) and multi-scale structural similarity (MS-SSIM)~\cite{ms-ssim}. Each dataset is split into training and validation sets with an 80\%:20\% ratio. The model is trained from scratch for 80 epochs using the Adam optimizer~\cite{adam} with a batch size of 8. The initial learning rate is set to 0.001 and is reduced by a factor of 0.1 if the validation metric does not improve for 8 epochs. Training is implemented in PyTorch and performed on an NVIDIA RTX~4090 GPU workstation. 

\textbf{DL model performance on RealGPR B-scans.} We evaluate the performance of the Sim-Net  and the FFDT-Net on RealGPR B-scans collected over a sand testbed. The experimental setup is shown in Fig.~\ref{fig:sandboxBscan}. A ceiling-mounted robotic arm moves the RealGPR along a predefined scan line to acquire B-scans above the testbed.  The testbed is filled with dry sand and small gravel and has RF absorbers attached to the inner walls to suppress boundary reflections. A metallic rod is buried at two depths: shallow (approximately 5~cm) and deep (approximately 15~cm). B-scans are collected with antenna-to-surface distances of approximately 15~cm and 25~cm over a 0.8~m scan trace with a  4~cm step size. The measured B-scans from 0 to 8~ns are preprocessed to remove $H_a$ and  are resized to 36$\times$819 before being fed into the model. The rescaled size is selected to match the time and spatial resolution of the training data.

The outputs of the two models for the measured B-scans are compared in Fig.~\ref{fig:network results}. For shallow-buried object cases, where the object reflection is strongly mixed with the sand-surface reflection, the Sim-Net removes most of the surface reflection clutter together with the overlapping object reflection. In contrast, the FFDT-Net removes background clutter while successfully recovering the object reflection. For deep-buried object cases, where the object reflection and surface clutter are more separated, the Sim-Net leaves noticeable residual clutter and does not fully restore the target reflection. In contrast, the FFDT-Net reliably removes background clutter while preserving the object reflection. Overall, compared to the Sim-Net, the FFDT-Net achieves substantially improved performance in  removing clutter and restoring object reflections. This superior performance is attributed to the improved training domain alignment enabled by FFDT. 

These experiments demonstrate that FFDT enables the construction of RealGPR-domain datasets from simulated data for training DL models. Compared with prior work, no \textcolor{black}{task-specific} RealGPR field measurements are required for model training or fine-tuning.

\section{\textcolor{black}{Discussion}}\label{sec5}

This section discusses the applicability and limitations of FFDT in practical applications. Specifically, Section~V-A analyzes the applicability and limitations of the dominant-reflection approximation used for $\theta$ selection; Section~V-B examines the sensitivity of FFDT to angle-estimation errors; \textcolor{black}{Section~V-C discusses the applicability of FFDT to different GPR systems;} Section~V-D discusses the use of FFDT to support learning-based methods for different GPR tasks; and Section~V-E discusses FFDT-based dataset generation for real-world sites with unknown and spatially variable soil permittivity.

\subsection{\textcolor{black}{Applicability and Limitations of the Dominant-Reflection Approximation in FFDT Implementation}}

\begin{figure}[tb]
	\begin{center}
		\begin{tabular}{c@{ }}
			\includegraphics[width=1\linewidth]{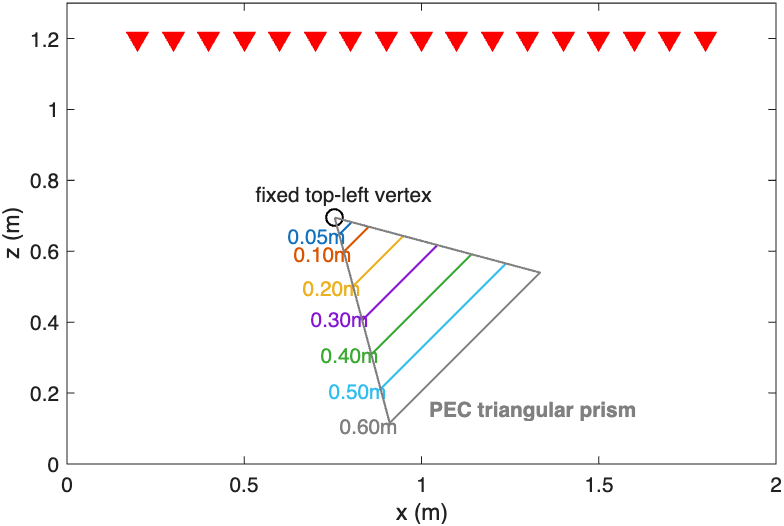}
			\footnotesize{(a)} \\
			\includegraphics[width=1\linewidth]{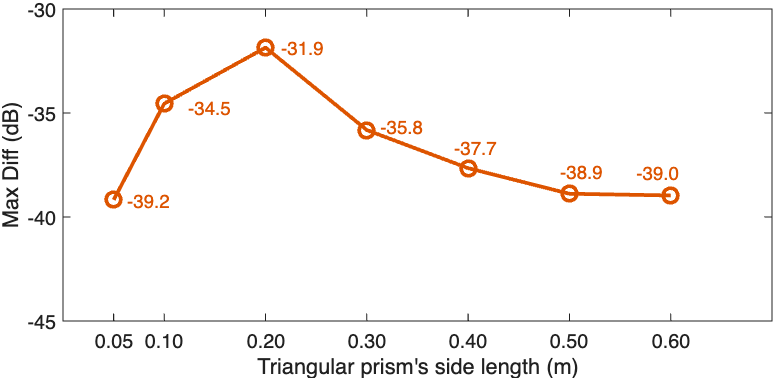}
			\footnotesize{(b)}  \\
		\end{tabular}
	\end{center}
	\caption{\textcolor{black}{(a) Cross-sectional view of the GPR measurement for PEC triangular prisms at 45$^\circ$ with different side lengths. (b) Performance of the FFDT using the dominant-reflection method to select $\theta$, quantified by the maximum Diff values between FFDT-transferred B-scans and the RealGPR B-scans.}}
    \vspace{-4pt}

	\label{fig:TriangleSize}
\end{figure}

\textcolor{black}{For targets with different geometries, the FFDT implementation in this work uses the dominant-reflection approximation to select the transmitting/receiving angle $\theta$. This approximation assumes that, at each survey position, the received target response is mainly governed by one dominant scattering component, or by several components distributed within a narrow angular range where the SimGPR and RealGPR radiation patterns vary only slightly. Under these conditions, the angle-dependent transfer function associated with the dominant reflection can also well represent the other major reflected components.  The approximation becomes less accurate when multiple strong scattering components with substantially different propagation angles contribute comparably to the received signal. In this case, applying a single $\theta$ inevitably introduces transfer-function mismatch for secondary components, leading to residual magnitude and phase errors. This behavior explains the increased Diff values observed for the rectangular prism and the 45$^\circ$-rotated triangular prism in Table~\ref{tab:peak_diff}, where strong reflections can originate from both faces and edges.}

\textcolor{black}{To further demonstrate the applicability and limitations of FFDT with the dominant-reflection approximation, FFDT performance is evaluated for a series of PEC triangular prisms at 45$^\circ$ with different side lengths, as shown in Fig.~\ref{fig:TriangleSize}(a). The targets share the same upper-left vertex, while the side length increases from 0.05 m to 0.6 m. This configuration allows the angular spread of the face and edge scattering responses to be gradually increased. Other simulation settings are the same as those in the case study. The maximum Diff between the FFDT-transferred B-scan and the RealGPR B-scan for different triangle sizes is summarized in Fig.~\ref{fig:TriangleSize}(b). For small targets, FFDT achieves good performance, with Diff around $-$40 dB, because the main scattering components occupy a very narrow angular range and the corresponding transfer functions are similar. As the target size increases, the Diff value first increases and then decreases. This trend results from the combined effects of the angular spread of the scattering components and the antenna transfer-function differences. For example, when the prism side length is 0.2 m, the face and edge reflections have comparable magnitudes in the received signal at some survey positions. Therefore, using only the face reflection for angle estimation introduces transfer-function mismatch for the edge-reflection components, causing the Diff to increase to around $-$30 dB. For larger targets, one reflection component tends to become dominant at a given scan position, whereas the other scattering components have longer propagation distances to the antenna and therefore smaller received magnitudes due to spherical spreading of waves. As a result, their influence on the final FFDT-transferred response becomes weaker, and the FFDT performance improves, with the Diff returning to around $-40$ dB.}

\begin{figure}[t]
	\begin{center}
			\includegraphics[width=0.75\linewidth]{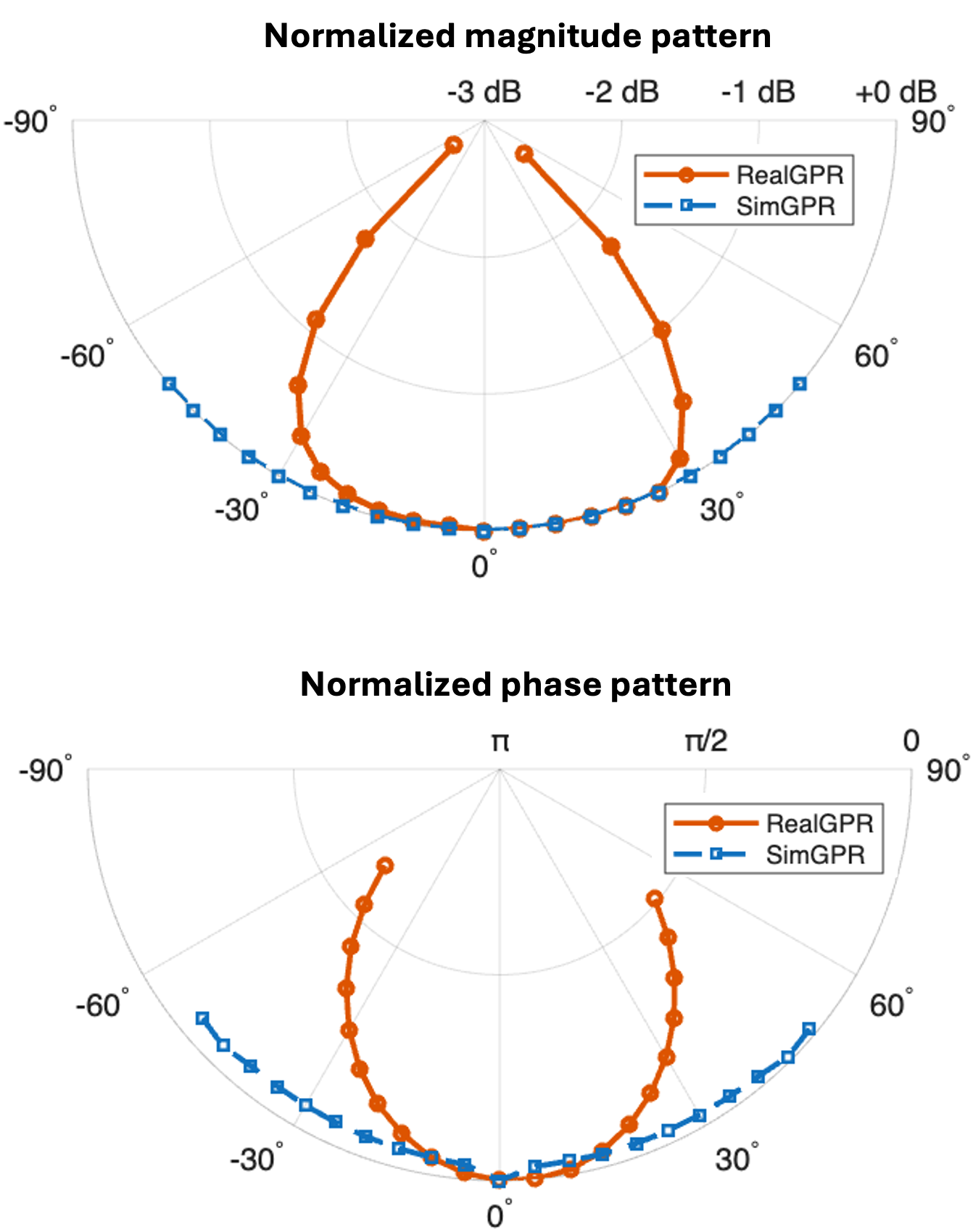}
            \vspace{-8pt}
	\end{center}
	\caption{\textcolor{black}{Normalized magnitude and phase of the radiation pattern of the SimGPR and RealGPR antennas in the experimental case in Section~\ref{MeasurementDemo1}.}}
    \vspace{-10pt}
	\label{fig:AntennaPatternComparison}
\end{figure}
\begin{figure}[tb]
	\begin{center}
			\includegraphics[width=1\linewidth]{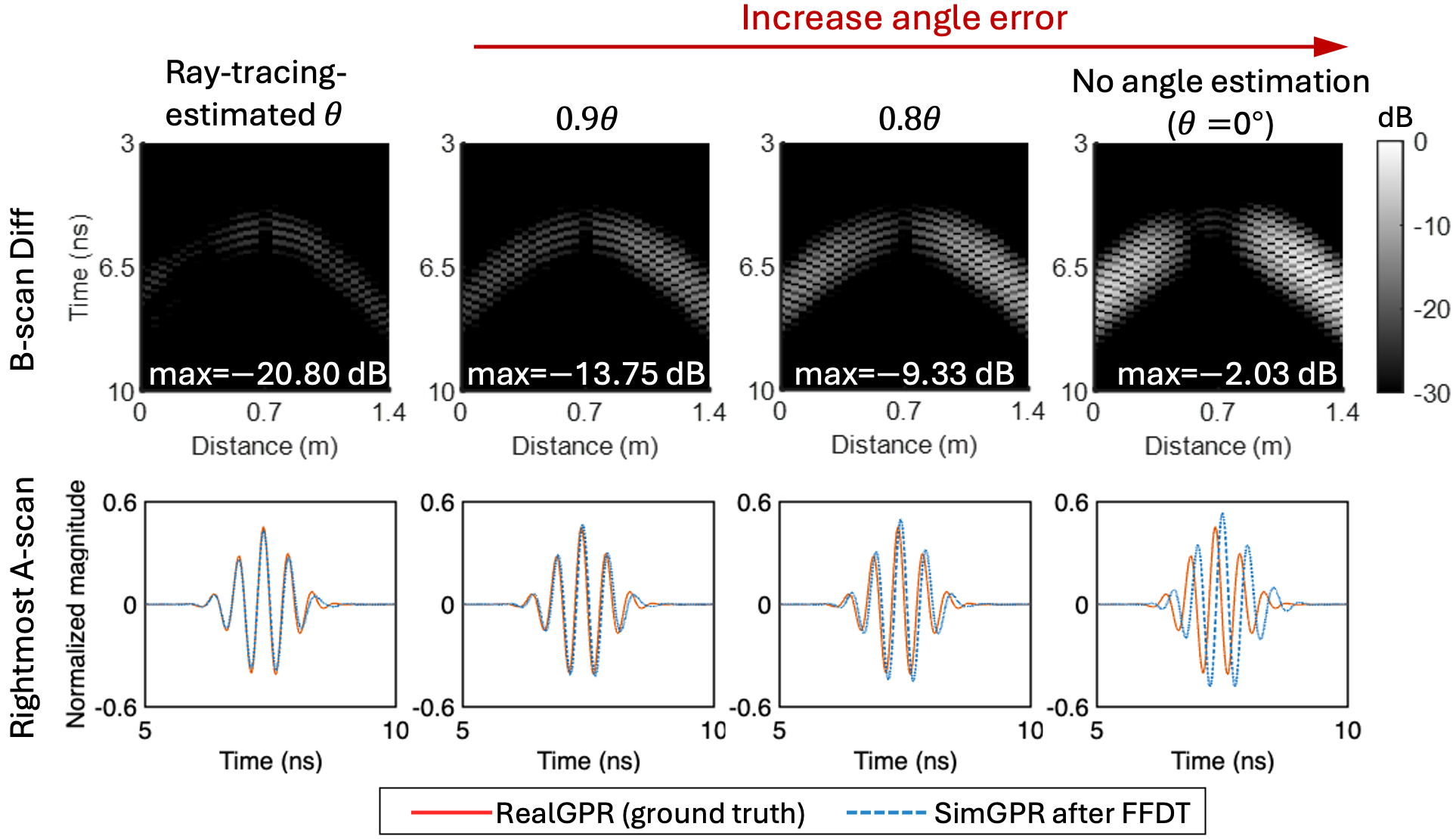}
            \vspace{-8pt}
	\end{center}
	\caption{\textcolor{black}{Comparison of FFDT performance for the scenario in Fig. \ref{fig:experiment setup example} under four angle implementations: ray-tracing-estimated angle $\theta$, angles underestimated to 0.9$\theta$ and 0.8$\theta$, and no angle estimation, where $\theta=0^\circ$. }}
    \vspace{-10pt}
	\label{fig:AngleErrorComp}
\end{figure}

\subsection{\textcolor{black}{Sensitivity of FFDT to Angle-Estimation Error}}
\textcolor{black}{A key step in FFDT is to identify the transmitting/receiving angle of each reflection component so that the corresponding transfer functions can be applied. Since these angle-dependent transfer functions model the angular radiation characteristics of the GPR antenna, the influence of angle-estimation errors on the FFDT performance depends on the radiation pattern difference between the SimGPR and RealGPR antennas. If the two antennas have similar radiation patterns, small angle-estimation errors are expected to have limited impact; in contrast, if they have substantially different angular magnitude or phase patterns, inaccurate angle estimation can introduce large transfer errors due to transfer-function mismatch.}

\textcolor{black} {To demonstrate the influence of angle-estimation errors, a comparative study is conducted using the metallic-rod experiment in Section~\ref{MeasurementDemo1}.  In this experiment, the SimGPR system uses a closely spaced Hertzian dipole and field probe as TX and RX, whereas the RealGPR system uses a monostatic Vivaldi antenna. Fig.~\ref{fig:AntennaPatternComparison} shows the normalized magnitude and phase radiation patterns of the two antenna setups. The clear pattern differences indicate that applying transfer functions at accurate angles is necessary for accurate sim-to-real transfer. FFDT performance is compared using four angle implementations: ray-tracing-estimated angles $\theta$, angles underestimated to 0.9$\theta$ and 0.8$\theta$, and an angle-independent case in which $\theta$ takes the downward-looking angle 0$^\circ$. The differences between the FFDT-transferred B-scans and the RealGPR B-scan are shown in  Fig.~\ref{fig:AngleErrorComp}. The corresponding maximum Diff values are $-$20.80, $-$13.75, $-$9.33, and $-$2.03 dB, respectively, which shows that FFDT performance gradually degrades as the angle-estimation error increases. The rightmost A-scan comparisons show increasing magnitude mismatch and time shift between the transferred and RealGPR A-scans. This demonstrates that applying transfer functions at inaccurate angles results in both magnitude and phase mismatches.}

\textcolor{black}{These results confirm that FFDT can be sensitive to angle-estimation errors when the SimGPR and RealGPR antennas have different angular patterns. However, since FFDT is designed for sim-to-real transfer from modeled simulation scenarios, where the medium properties, object positions and characteristics are known, ray tracing and dominant-reflection selection can be used to reduce angle-estimation errors and support effective FFDT implementation.}

\begin{table}[tb]
\centering
{\color{black}
\caption{Target Detection and Classification Performance}
\label{tab:yolo_results}
\begin{tabular}{lccc}
\hline
\hline
Training dataset & $\mathrm{mAP}_{50}$ & $\mathrm{mAP}_{70}$ & $\mathrm{mAP}_{90}$ \\
\hline
FFDT-transferred dataset & 0.9941 & 0.8792 & 0.5899 \\
SimGPR dataset           & 0.8996 & 0.7683 & 0.1828 \\
\hline
\hline
\end{tabular}
}
\end{table}

\subsection{\textcolor{black}{Applicability of FFDT to Different GPR Systems}}
\textcolor{black}{GPR systems can differ substantially in antenna type, front-end circuitry, time- or frequency-domain acquisition scheme, and operating frequency. For FFDT operation, differences in antenna, shielding, and front-end response are incorporated through the system transfer functions defined in Section II-A. FFDT operates in the frequency domain over the common frequency band of the SimGPR and RealGPR systems. Therefore, its implementation does not depend on whether the GPR data are originally acquired in the time or frequency domain, as time-domain data can be transformed into the frequency domain using the Fourier transform.}

\textcolor{black}{The current FFDT formulation is applicable to monostatic and quasi-monostatic, co-polarized GPR systems operating under far-field conditions. For quasi-monostatic systems, the TX-RX separation should be sufficiently small relative to the sensing distance such that the bistatic angle is small and the TX and RX propagation paths can be approximated by the corresponding monostatic path. Experimental studies of monostatic-to-bistatic equivalence have shown that, for complex targets, the monostatic approximation can remain reasonably accurate for bistatic angles of approximately  5$^\circ$-10$^\circ$ \cite{BistaticScattering}. Extending the framework to fully polarimetric, strongly bistatic, and near-field configurations represents important research directions to broaden the applicability of GPR domain transfer methods.}

\textcolor{black}{From a practical implementation perspective, FFDT for commercial GPR systems requires controlled system calibration, in which the antenna-to-calibration-target distance and orientation should be accurately controlled using a positioning stage, robotic arm, or similar setup. A practical approach is to perform this one-time system characterization during system manufacturing, when controlled positioning equipment and calibration environments are more readily available. The resulting system-specific transfer functions can then be stored and reused by end users for subsequent FFDT applications, provided that the system configuration remains unchanged. We also welcome collaboration with GPR manufacturers and users to support system calibration for broad FFDT implementation.}

\subsection{\textcolor{black}{Applicability of FFDT to Different GPR Tasks}}

\textcolor{black}{Considering FFDT’s ability to transfer SimGPR signals to RealGPR signals with accurately matched time-domain pulse position, waveform shape, magnitude, and phase, FFDT can benefit broader GPR tasks beyond the clutter-removal task demonstrated in Section~\ref{MeasurementDemo2}, such as object detection, classification, and imaging. }

\textcolor{black}{As a demonstration, we conduct a complementary numerical study using YOLO11 model~\cite{YOLOv11} for joint underground-target detection and classification. Motivated by the task considered in DDA-GPR~\cite{UnsupervisedLearningReference1}, three urban-road target classes are considered: cavities, manholes, and pipelines. The SimGPR and RealGPR systems implemented in this task are the two different gprMax-modeled systems used in the case study in Section~\ref{sec:CaseStudy}. The dataset is generated following the heterogeneous road–subsurface model and material properties in \cite{UnsupervisedLearningReference1}, \cite{DDA_model}. A total of 480 scenarios are simulated with SimGPR, with 160 scenarios for each target class. Each original B-scan is cropped into five overlapping sub-B-scans with a lateral span of 2.0 m, resulting in 2,400 samples. The SimGPR B-scans obtained after removing $H_{a}$ constitute the SimGPR dataset, and applying FFDT produces the corresponding FFDT-transferred dataset. Each target in the B-scans is annotated with its class and a bounding box covering its reflection response for network training. The YOLO11 model is trained separately on the SimGPR and FFDT-transferred datasets using identical settings.}

\textcolor{black}{The trained models are evaluated on 45 RealGPR B-scans, with 15 cases for each target class. Model performance is evaluated using mean average precision (mAP)~\cite{mAP} at intersection-over-union (IoU) thresholds of 0.50, 0.70, and 0.90, denoted as $\mathrm{mAP}_{50}$, $\mathrm{mAP}_{70}$, and $\mathrm{mAP}_{90}$, respectively. As shown in Table~\ref{tab:yolo_results}, the model trained on the FFDT-transferred dataset achieves higher mAP values than the model trained on the SimGPR dataset at all three IoU thresholds. The improvement is most pronounced for $\mathrm{mAP}_{90}$, where the predicted bounding box must closely match the ground-truth box. These results demonstrate that the FFDT-transferred dataset improves target detection and classification performance and further prove that FFDT can benefit learning-based methods for other GPR tasks.}

\subsection{\textcolor{black}{FFDT-Based Dataset Generation for Real-World Sites}} \label{apply to practical field}

\textcolor{black}{When FFDT is used to generate training datasets for a specific site, the actual soil permittivity may be unknown and spatially variable. In such cases, the following strategy can be adopted. First, the approximate range of bulk soil permittivity at the site can be estimated from prior knowledge of the soil or auxiliary measurements. For example, different soil textures and moisture ranges can be converted to soil permittivity ranges using the Peplinski mixing model~\cite{Peplinski}; dielectric probes or common-midpoint GPR measurements can also be used to estimate site-specific permittivity ranges. Second, instead of constructing simulation scenarios using a single soil permittivity, many scenarios can be built to cover the estimated range of soil permittivity and heterogeneity. FFDT can then be applied to the simulated data from these scenarios to generate a diverse RealGPR-domain dataset that incorporates site-specific soil-permittivity variability. This strategy treats variability in soil permittivity as part of the simulation variability. By covering site-specific variations in soil permittivity and heterogeneity, the resulting FFDT-transferred dataset supports the training of more robust downstream learning-based models for the target site.}

\section{\textcolor{black}{Comparison with Related Methods}}\label{sec6}

\subsection{\textcolor{black}{Comparison with GPR Calibration Methods}}

\begin{figure*}[tb]
	\begin{center}
			\includegraphics[width=0.95\linewidth]{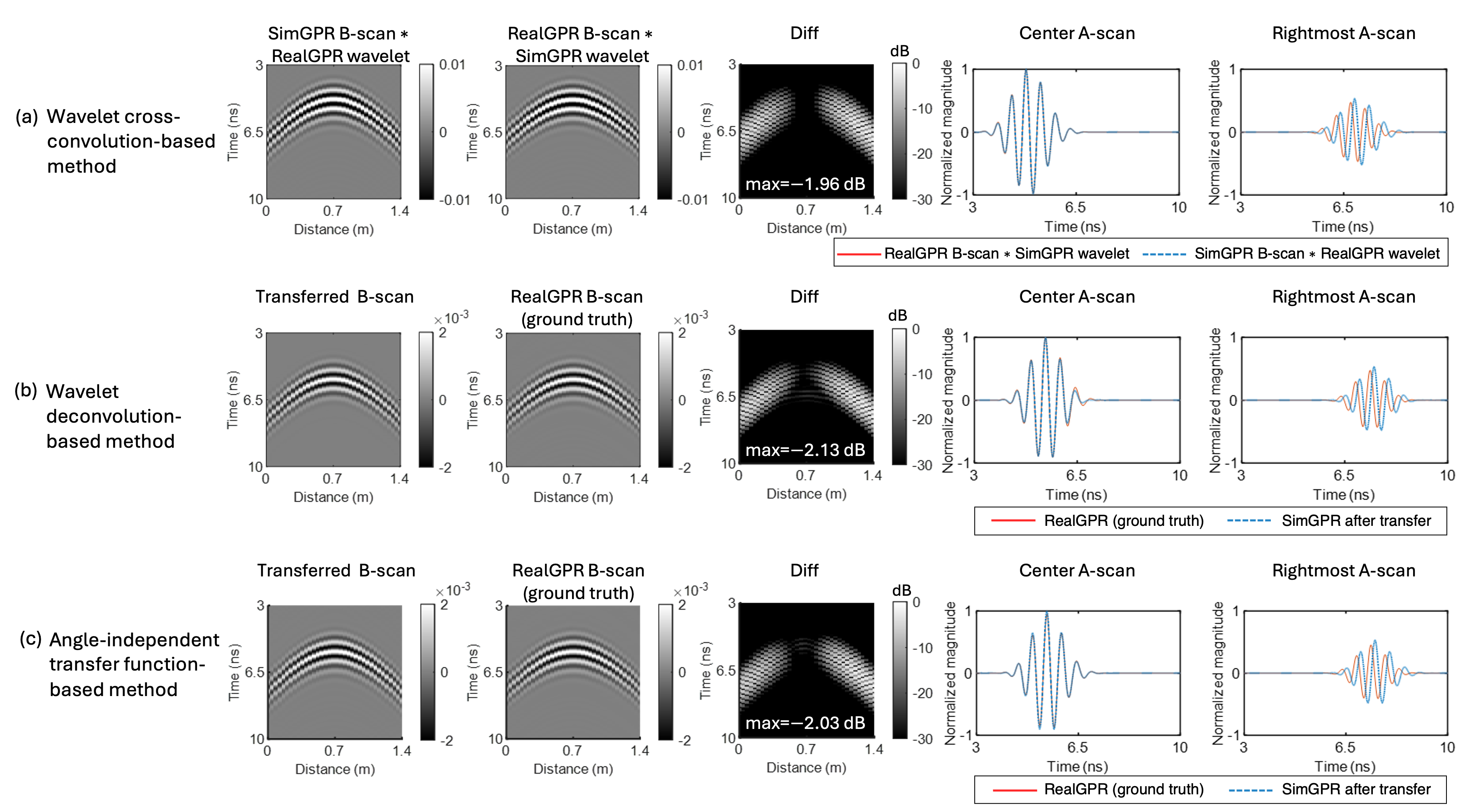}
            \vspace{-8pt}
	\end{center}
	\caption{\textcolor{black}{Sim-to-real domain-transfer performance using representative calibration concepts: (a) wavelet cross-convolution-based method, (b) wavelet deconvolution-based method, (c) angle-independent transfer function-based method.
}}
    \vspace{-0pt}
	\label{Fig:Methodcomparison_FWI_SBD_TransferFunction}
\end{figure*}

\textcolor{black}{FWI is an inversion method for estimating subsurface electromagnetic properties by iteratively updating the subsurface model and minimize the mismatch between simulated and measured GPR responses \cite{FWI_crossconv2, Feng2023DataEncoding}. Different calibration techniques can be incorporated into or used alongside FWI to reduce mismatches associated with the GPR source waveform and system response \cite{FWI_crossconv1,FWI_crossconv2,FWI_crossconv3,FWI_deconv,SBD1,SBD2,Lambot1,lambot3}. In contrast, FFDT neither inverts measured data nor updates the subsurface model; instead, it is developed for sim-to-real data generation and domain alignment. Given a known simulated scenario, FFDT uses angle-dependent system transfer functions to map SimGPR signals into the signal domain of a specified RealGPR system. FFDT can therefore complement FWI and other inversion methods by improving the consistency between simulated and measured GPR signal domains.}

\textcolor{black}{To quantitatively compare FFDT with common calibration methods implemented in FWI, we adapted these representative calibration methods to the sim-to-real transfer task in the metallic-rod experiment in Section~\ref{MeasurementDemo1}. The wavelet cross-convolution method~\cite{FWI_crossconv1,FWI_crossconv2,FWI_crossconv3} is implemented by convolving the SimGPR B-scan with a RealGPR reference wavelet and convolving the RealGPR B-scan with a SimGPR reference wavelet. The two cross-convolved B-scans are compared in Fig.~\ref{Fig:Methodcomparison_FWI_SBD_TransferFunction}(a). The wavelet deconvolution calibration method, which represents iterative wavelet-estimation and sparse blind-deconvolution approaches~\cite{FWI_deconv,SBD1,SBD2}, is implemented by first deconvolving the SimGPR B-scan using an effective SimGPR wavelet and then reconvolving the result with an effective RealGPR wavelet. The transferred B-scan and the RealGPR B-scan are compared in Fig.~\ref{Fig:Methodcomparison_FWI_SBD_TransferFunction}(b). The angle-independent transfer-function method~\cite{Lambot1,lambot3} is implemented by transferring all reflection components using the downward-looking transfer function. This implementation can be regarded as a degraded version of FFDT without angular dependence. The transferred B-scan and the RealGPR B-scan are compared in Fig.~\ref{Fig:Methodcomparison_FWI_SBD_TransferFunction}(c).}

\textcolor{black}{As shown in Fig.~\ref{Fig:Methodcomparison_FWI_SBD_TransferFunction}, these calibration methods effectively reduce waveform mismatch in the downward-looking direction. However, as shown in the rightmost A-scan comparisons, residual magnitude and waveform-delay differences remain when target reflections arrive from nonzero propagation angles. This is because the reference wavelet or transfer function corresponds to only the downlooking direction and does not account for angle-dependent wavelet or transfer-function differences. As a result, the maximum differences between the transferred B-scans and the RealGPR B-scans for these three methods are $-$1.96 dB, $-$2.13 dB, and $-$2.03 dB, respectively. In contrast, by applying calibrated angle-dependent transfer functions at the estimated propagation angles, FFDT reduces the maximum difference to $-$20.80 dB, as shown in Fig.~\ref{fig:sim2realDIFF}. This comparison demonstrates that, for sim-to-real transfer with directional GPR antennas, combining FFDT with an angle-dependent GPR model enables more accurate domain transfer.}

\subsection{\textcolor{black}{Comparison with Unsupervised Learning-Based Domain Adaptation and Data Augmentation Methods}}

\textcolor{black}{Unsupervised learning-based methods have been developed to reduce the sim-to-real domain gap, including \textcolor{black}{adversarial domain adaptation methods that learn domain-invariant feature representations from labeled SimGPR data and unlabeled task-specific RealGPR data to improve model generalization across domains 
~\cite{UnsupervisedLearningReference1,UnsupervisedLearningReference2,UnsupervisedLearningReference3}}, and generative adversarial network (GAN)-based methods that generate RealGPR-like B-scans for data augmentation~\cite{DataAugmentation1,DataAugmentation2}. These methods are effective for improving task-specific model performance, but they still require substantial task-specific RealGPR data during model training or adaptation.}

\textcolor{black}{In contrast, FFDT is a physics-based domain-transfer method that directly map data from the SimGPR domain to the RealGPR domain.} After one-time SimGPR and RealGPR system calibration, FFDT can generate RealGPR-domain B-scans from new simulated scenarios without requiring task-specific RealGPR field measurements. Moreover, unlike GAN-generated images, the FFDT-transferred B-scans are physically paired with the corresponding simulated ground truth, including quantitative subsurface permittivity distribution that defines the target geometry, size, and material properties. Therefore, FFDT is particularly useful for generating paired RealGPR-domain datasets to support both supervised and unsupervised learning methods for broader GPR tasks such as clutter removal, target detection, parameter estimation, and quantitative imaging.

\section{Conclusion}\label{sec7}

A physics-based far-field GPR domain transfer method (FFDT) is presented to map SimGPR data to corresponding RealGPR-domain data over any sub-band within the overlapping frequency range of the two systems. The method is developed by modeling SimGPR and RealGPR using angle-dependent transfer functions that capture three-dimensional antenna radiation characteristics, and by integrating ray tracing to select and apply the appropriate transfer functions to transfer each subsurface reflection. Extensive studies across diverse object and subsurface medium scenarios have been conducted to validate FFDT performance. The difference between the FFDT-transferred data and the corresponding RealGPR ground truth is reduced to below $-$30~dB in most cases, demonstrating that FFDT effectively reduces the sim-to-real domain gap.
 
A primary application of FFDT is the generation of RealGPR-domain datasets from simulations for training DL models. This capability has been demonstrated using an experimental clutter-removal task and a numerical detection and classification task, where a DL model trained on a FFDT-transferred dataset achieves robust performance on RealGPR data.  By enabling RealGPR-domain dataset generation without requiring \textcolor{black}{task-specific} field measurements, FFDT addresses the long-standing bottleneck of RealGPR data scarcity and provides a practical foundation for developing and deploying  DL-based GPR processing methods in real-world applications. 

\textcolor{black}{FFDT is developed for time-invariant, (quasi-)monostatic, co-polarized GPR systems operating under far-field conditions. The current method is particularly suitable for elevated GPR configurations, such as drone- and vehicle-mounted systems for landmine detection ~\cite{LandmineFarField1,LandmineFarField2,LandmineFarField3}, soil moisture mapping~\cite{SoilFarField1,SoilFarField2,SoilFarField3}, and snow, ice, and glacier mapping \cite{GlacierFarField1,GlacierFarField2,GlacierFarField3,GlacierFarField4}. Future work will extend the framework to near-field, strongly bistatic, and fully polarimetric configurations to further broaden its applicability.}

\section*{Acknowledgments}
Hai-Han Sun gratefully acknowledges Professor Sébastien Lambot at Université Catholique de Louvain for insightful discussions on GPR technologies in 2024 and for his publications that inspired this work. The authors also thank the anonymous reviewers for their valuable and constructive feedback.

\bibliographystyle{IEEEtran}
\bibliography{refs}

\end{document}